# Digital Qiang Garden: Artist Residency as Interface for Creativity Under Change

**Yunyi Rainee Wang (yunyiraineewang@gmail.com) [1]**
**Joshua Nijiati Alimujiang (alimujiang.nijiati@polytechnique.edu) [2]**

**[1] Harvard Graduate School of Education, Cambridge, Massachusetts, United States**
**[2] École Polytechnique, Paris, France**

## ABSTRACT

This pictorial presents *Digital Qiang Garden*, an artist-led residency situated in Qiang ethnic minority villages in western Sichuan, China, as a practice-based inquiry into creativity for change. Rather than framing creativity as a discrete outcome or individual agency, the project examines how creative practice unfolds through lived conditions under uncertainty. Drawing from three years of residency practices, the pictorial identifies four creative modes of praxis that repeatedly emerged through shared living, material engagement, and cross-media collaboration. Through annotated visuals and project-based case studies, we further trace how these modes generated eight interrelated layers of change: spanning artistic processes, social relations, cultural encounters, ecological awareness, organizational structures, and artists' longer-term development trajectories. By framing an artist-led residency as a transferable structure, this work contributes to human-computer interaction and creativity research by foregrounding designed conditions, situated action, and relational processes through which change unfolds over time.

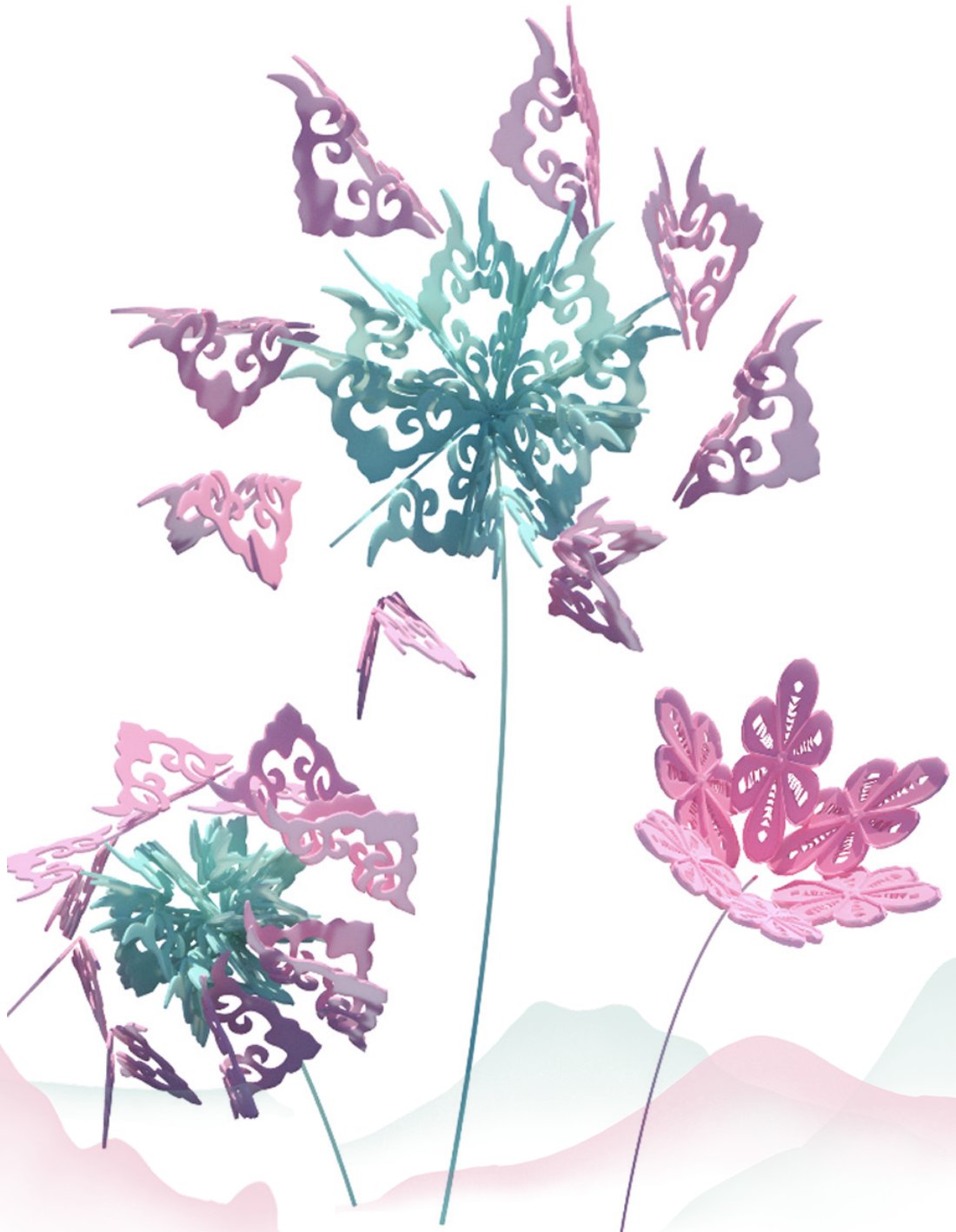

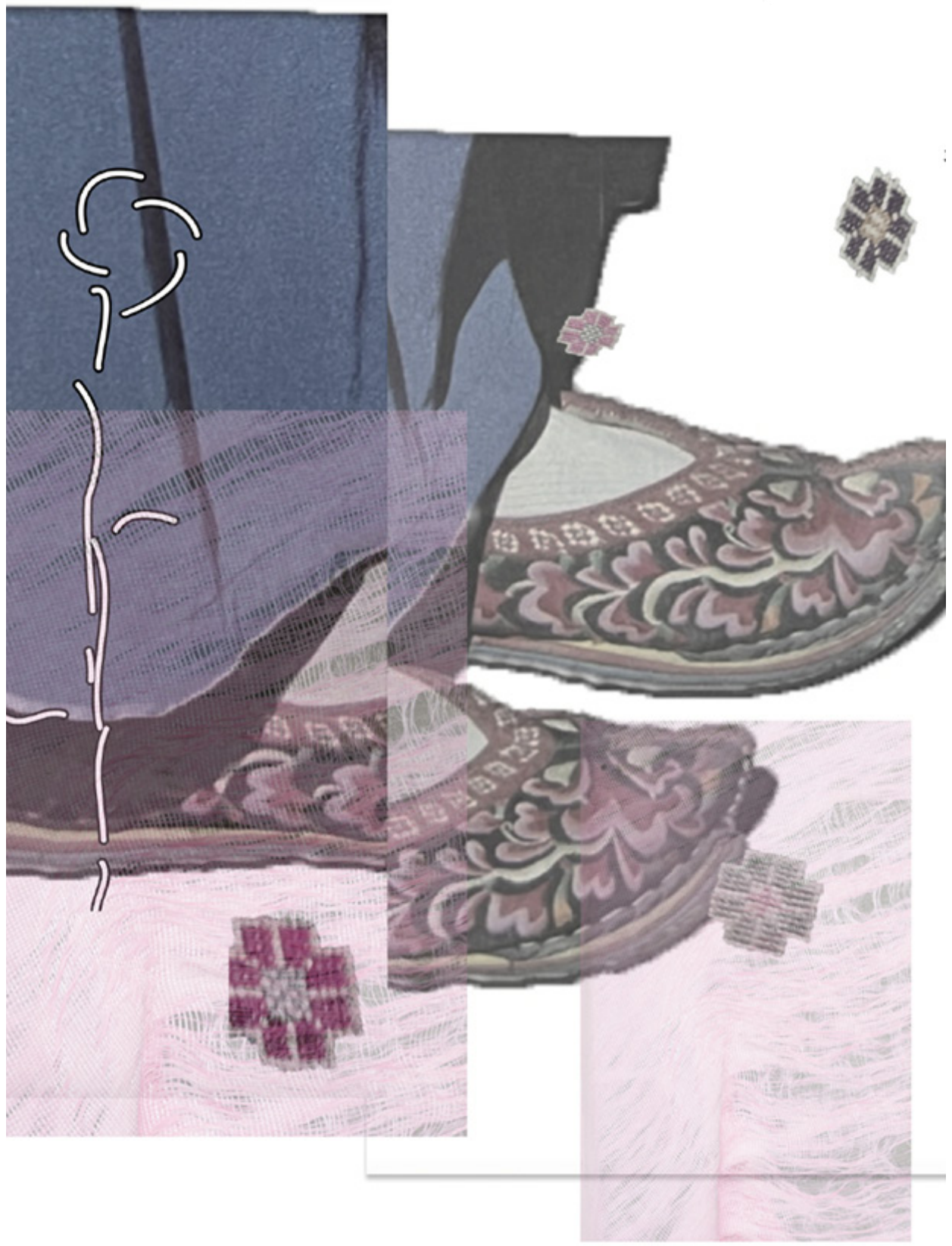

*Figure 1: Qiang people's traditional embroidery shoes*

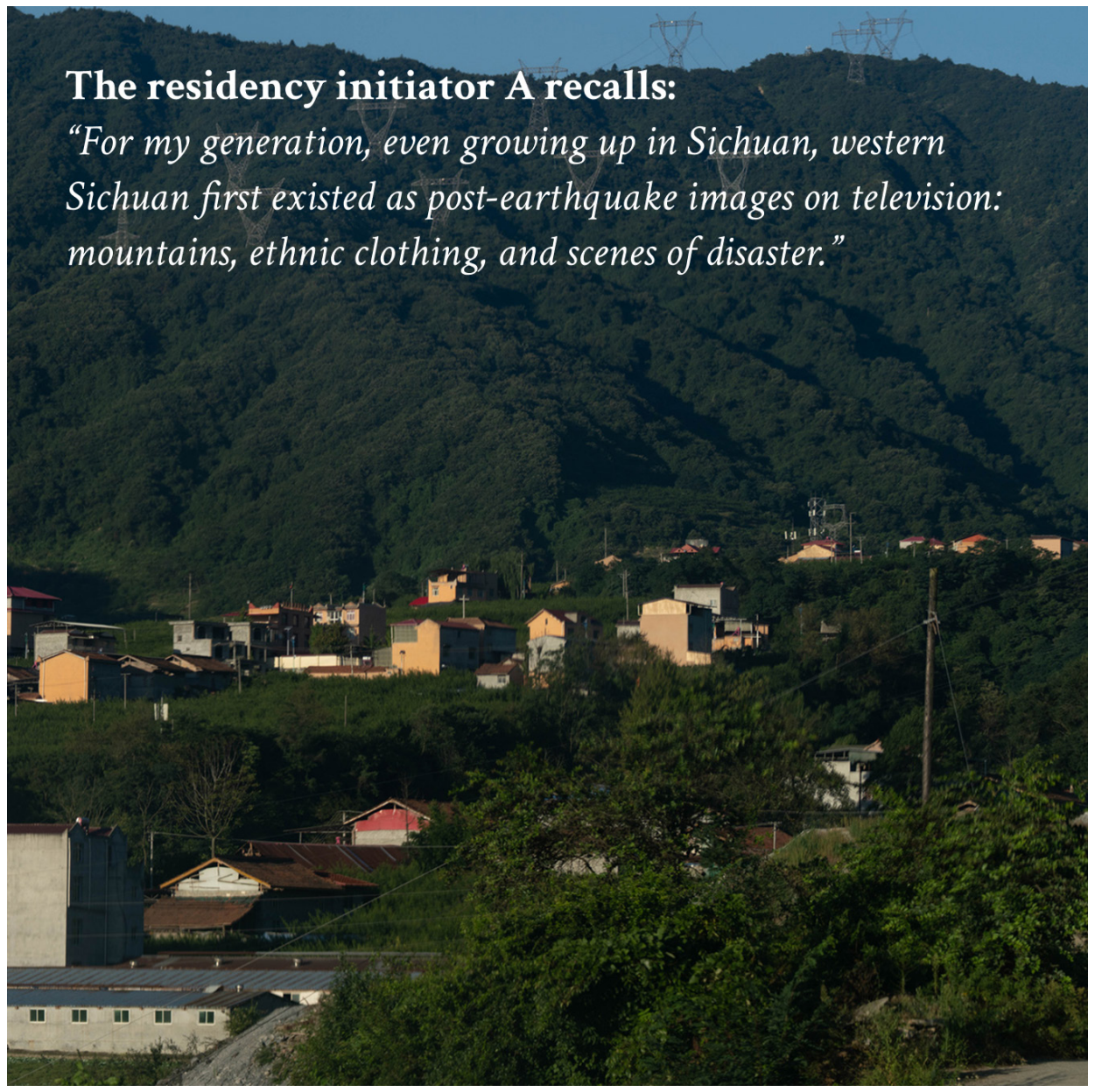

*Fig 2: Shang'guan village, Mao county, Sichuan*

## INTRODUCTION

### Place as Condition

Mao County is located in Aba Tibetan and Qiang Autonomous Prefecture, Sichuan, China, where the Qiang people, one of China's oldest ethnic minority groups, comprise over ninety percent of the population [1]. Following the 2008 Wenchuan earthquake, the mountainous regions of western Sichuan entered public imagination primarily through screens, framed by images of disaster, loss, and collective mourning [2].

In the decade that followed, post-disaster reconstruction, migration, and development policies profoundly reshaped village life across the region. While infrastructure was rebuilt, everyday practices were altered or displaced, including traditional housing, embodied craft knowledge, everyday rhythms, rituals, and festivals [2].

At the same time, ethnic aesthetics increasingly circulated as visual symbols, even as the practices and bearers sustaining them became fragile or endangered [3] [4].

Here, geography, climate, infrastructure, and cultural transmission are not neutral backdrops but active forces shaping daily life. Knowledge circulates through bodies, materials, gestures, and shared time more than through formal archives or institutions [5]. This condition calls for more embodied, sensorial, and tangible approaches to preserving culture, heritage, and community.

### From Encounter to Question

A return visit to Mao County after the residency initiators' graduation transformed this awareness from distant empathy into bodily experience, from artistic inspiration-seeking into an urgent call to action: Could artist-residency-based creativity contribute to engaging with living cultural practices, local economies, and everyday life, rather than observing them from a distance?

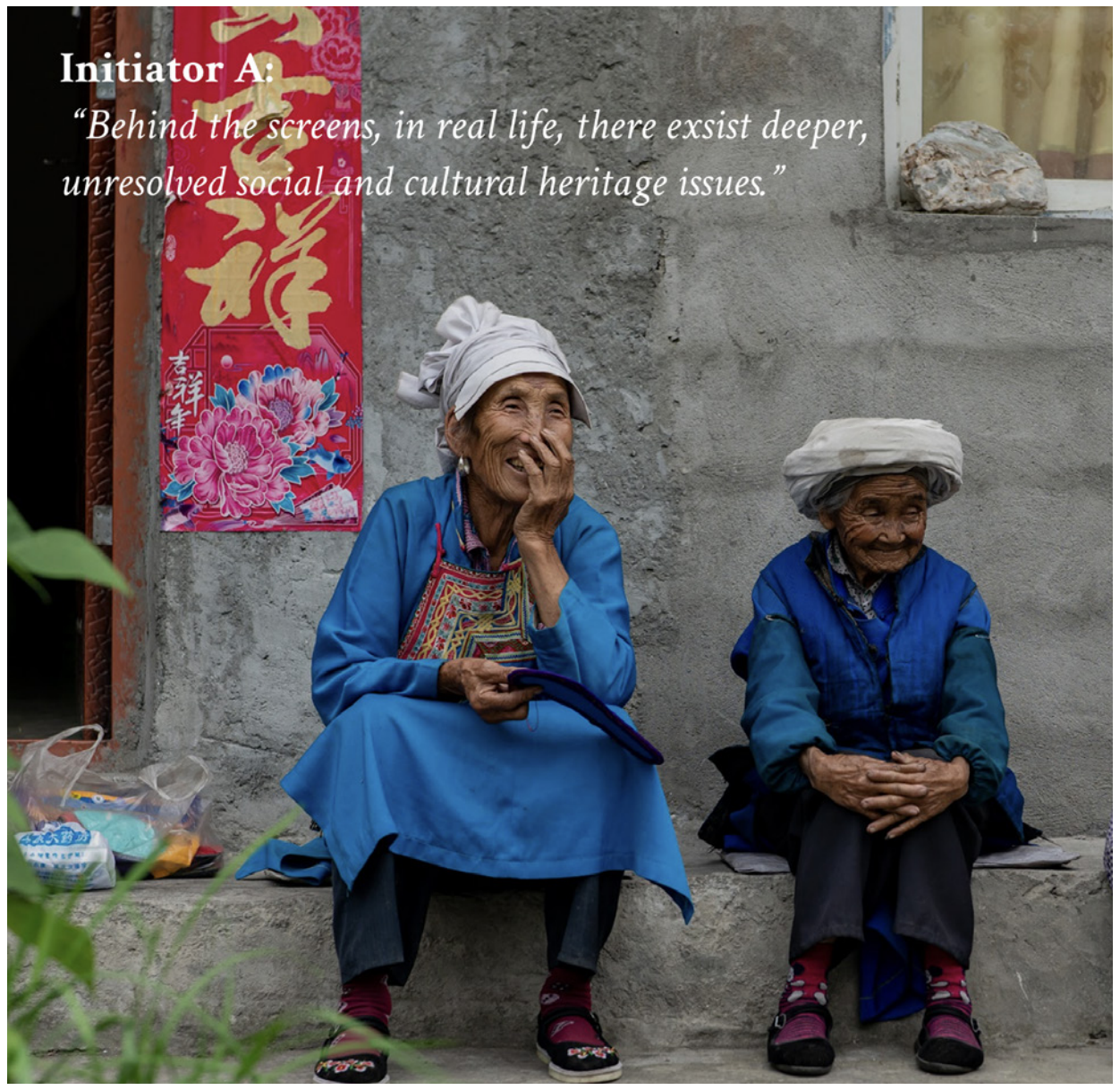

*Fig 3 : Senior Qiang women chatting with Us*

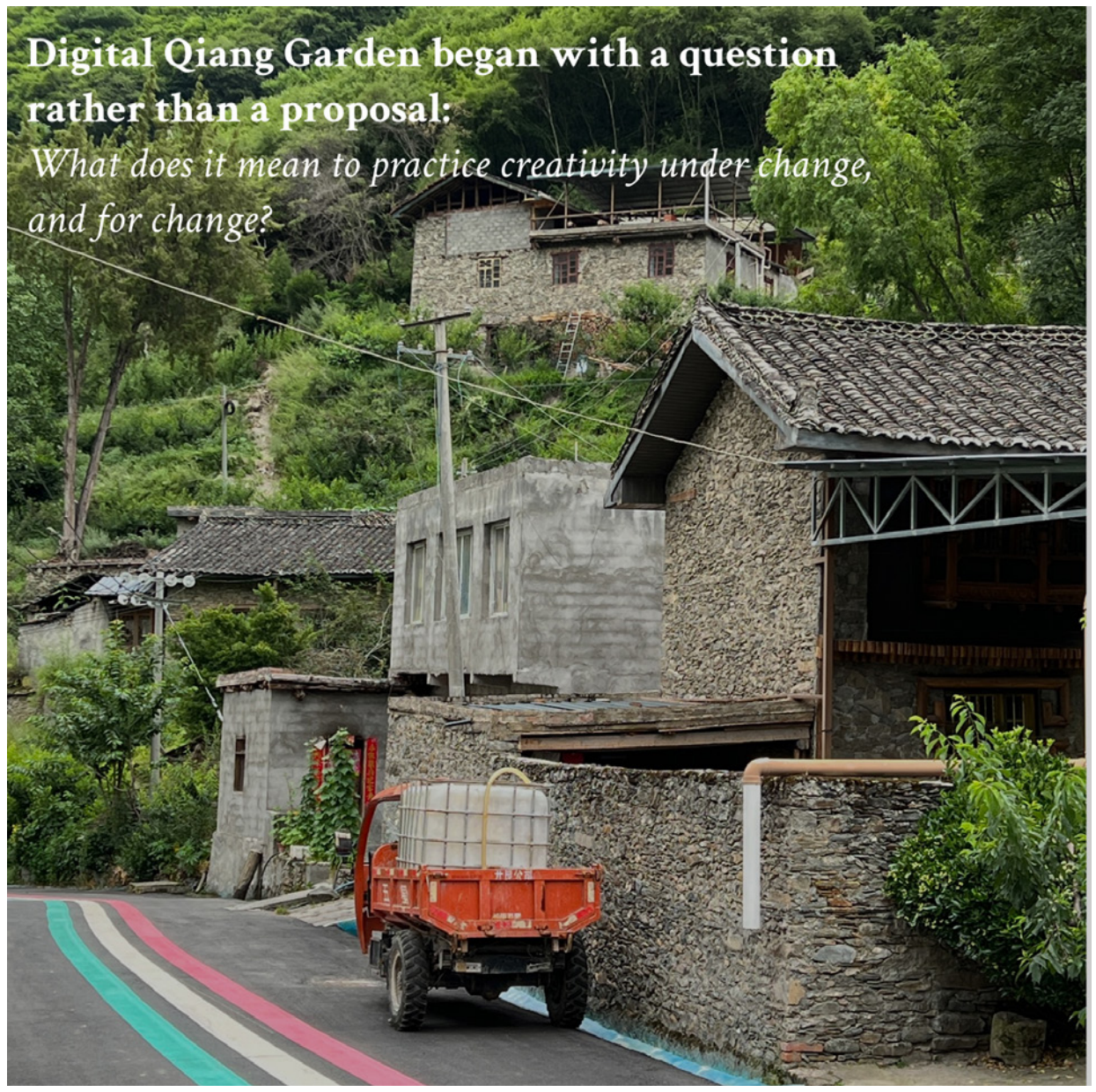

*Fig 4: Infrastructure in the Qiang village*

The question was no longer how to represent contexts through art, but how to act within them responsibly. In contexts shaped by post-disaster reconstruction, cultural vulnerability, and uneven development, producing a singular artwork felt insufficient, and potentially extractive. This challenge could not be addressed through solitary artistic practice alone. Instead, it called for a collective, interdisciplinary system, an interface that operates across media, cultures, and communities; an interface capable of navigating creativity as it unfolds within lived conditions, ambiguity, and uncertainty.

## WHAT IS DIGITAL QIANG GARDEN?

*Digital Qiang Garden* is an annual, artist-led residency initiated in 2023 by independent artist-researchers and operated as an international and interdisciplinary artist collective. It is situated in Qiang ethnic minority villages in Aba Tibetan and Qiang Autonomous Prefecture, Sichuan, China. Each edition brings together and fully sponsors a small cohort of artists and researchers (typically 6–12 participants) working across digital and new media art, design, architecture, music, film, emerging technologies and other creative practice. Participants live and work together in a rural village for approximately two weeks, sharing daily life alongside research, making, and exchange.

Rather than following a fixed theme or production brief, the residency is structured around open-ended inquiry shaped by local conditions. Creative work emerges through shared living, field engagement, daily encounters, and artistic experiments, responding to place and the local cultural through process rather than predetermined outcomes. To date, the collective has brought together 24 artists recruited globally and supported the development of over 30 projects rooted in Qiang cultural contexts. These works have since been exhibited and presented internationally.

Over three years, *Digital Qiang Garden* has evolved not as a fixed program, but as a recurring structure that adapts to changing sites, participants, and conditions.

**Digital - hybrid practice** **Qiang - lived culture** **Garden - collective care**

*An experiment by a small group of recently graduated artist-researchers.*
*A summer departure.*
*A self-built utopia.*
*A decision to act before fully knowing what could be changed.*

*What does it mean to practice creativity under change—and for change?*

## WHY ARTIST-LED RESIDENCY?

In Mao County, what an institutional program would need to import at great cost is already present, a place to live and work, materials in reach, and cultural knowledge carried through everyday gestures. To creatives, the village is not “underdeveloped,” it is differently resourced.

We built a residency because the change we care about does not happen in a studio. Creativity here is not an individual act, but a situated practice shaped by place, bodies, and relations [6].

*Digital Qiang Garden* adopts an artist-led structure to decentralize authorship and decision-making. Instead of competing for scarce opportunities, we assembled a lightweight structure through artists’ networks, mutual support, and a low budget spent locally. The residency itself is collectively built, negotiated, and sustained, allowing creative practice to remain responsive to place, people, and uncertainty.

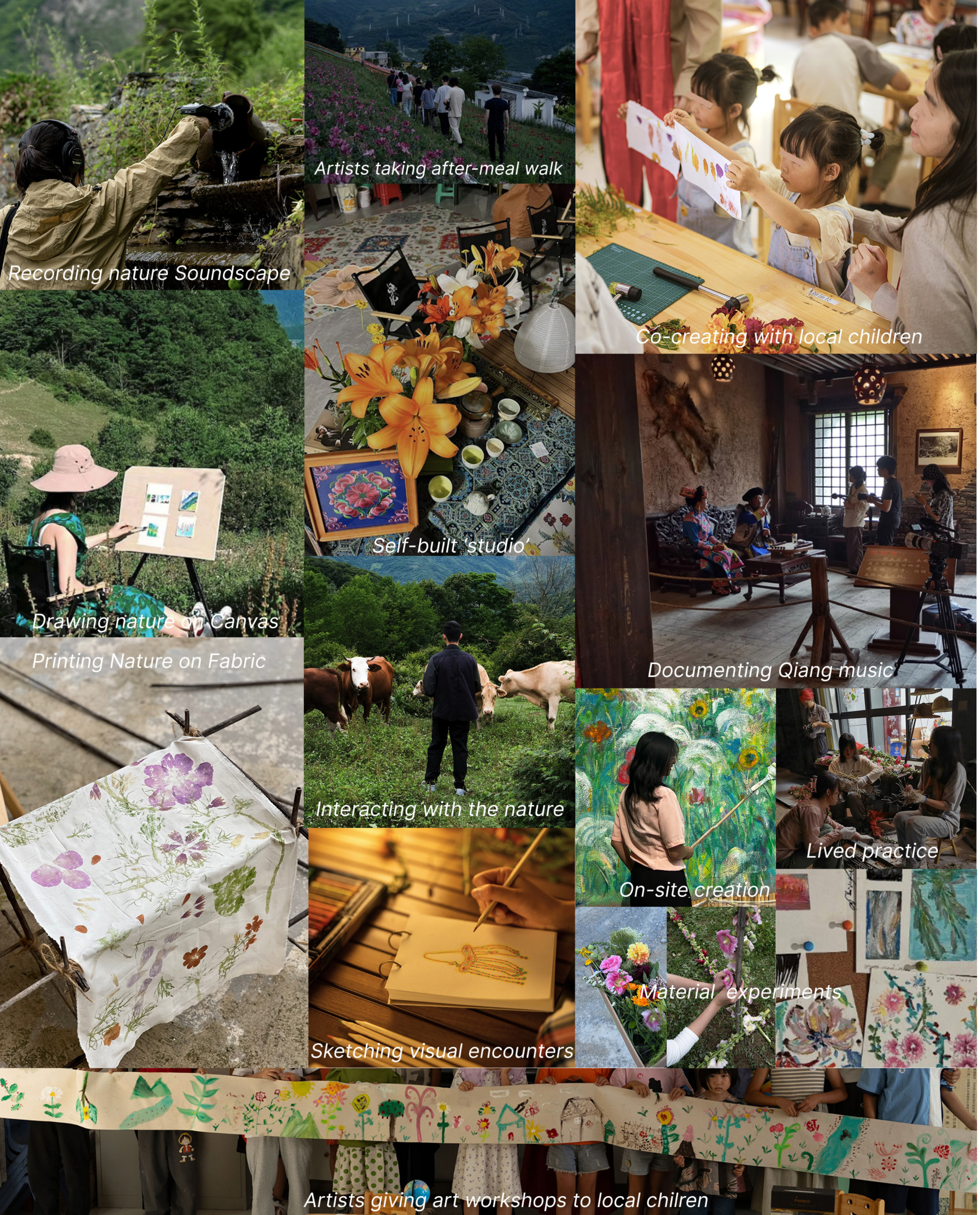

## METHODS

*Digital Qiang Garden* draws from research through design and practice-based inquiry in Human Computer Interaction (HCI), where knowledge is articulated through making, reflection, and documentation of situated work rather than controlled experiments [7][8]. It also builds on participatory design's shift from users, audience, community members to co-creative publics, treating participation as configured through relationships, responsibilities, and context [9][10][11].

Methods are enacted through shared living, field engagement, collective making, and critical reflecting as Figure 5 shown for a continuous iteration process. Material exploration, workshops, and documentation function simultaneously as process and outcome, foregrounding tacit, sensory, and relational ways of knowing [5][12][13]. This pictorial is co-authored by the residency initiator and a participating artist, holding together how conditions were built and how practice unfolded over time. Moreover, this pictorial holds both organizational and situated perspectives, reflecting on the conditions and rhythms that allow creative practice to adapt under change.

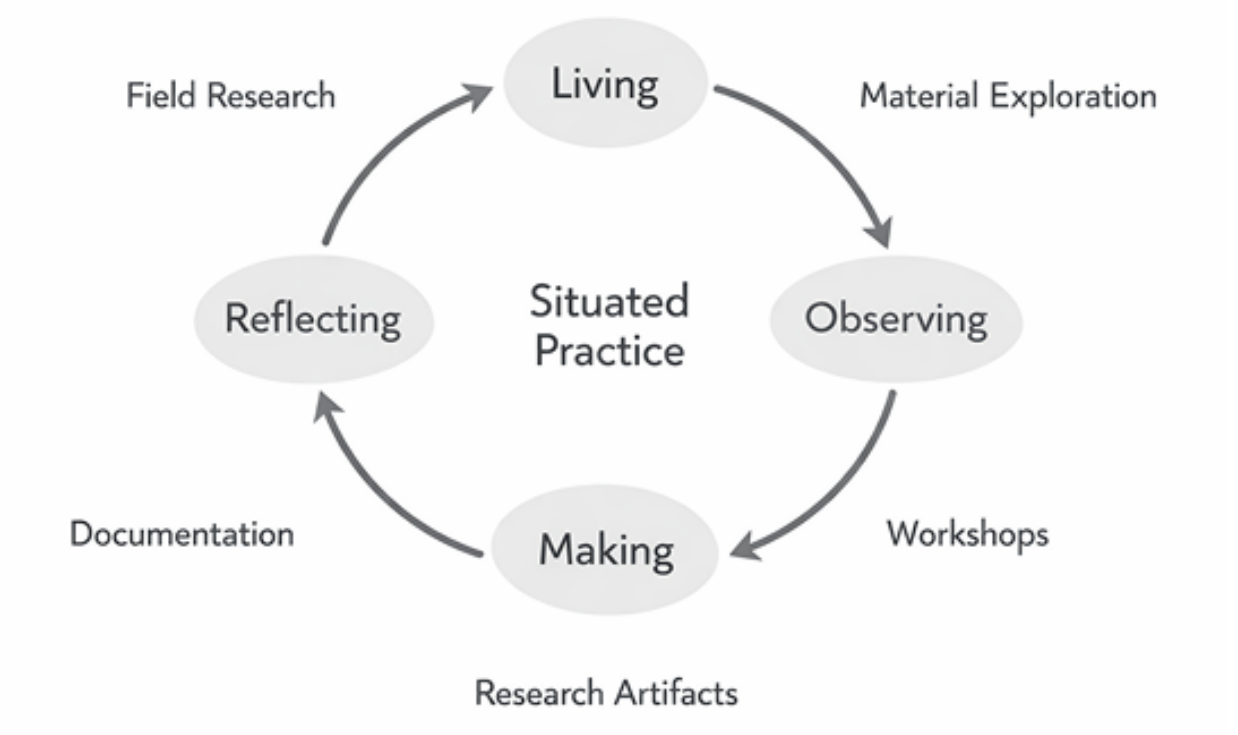

*Fig 5: Methods Enacted Through Practice*

**Heritage as Creative Ground**

The residency's creative thread is rooted in Qiang heritage. Instead of sampling it as a source of exotic motifs, we approach heritage as a living practice that challenges cultural displacement. We critique any practice that involves cultural appropriation by stripping symbols with only subjective conjecture for aesthetic consumption. Every practice in *Digital Qinag Garden* is rooted in interview, workshop, and co-creation with community members. Figure 7 to 10 maps key domains of Qiang cultural life that informed our fieldwork: ritual and belief practices, seasonal festivals, music and performance, and craft traditions [14]. The residency schedules access to heritage bearers through interviews, demonstrations, and shared time in situ. Co-creation then becomes a method: artists work with embodied knowledge through collaborative making with practitioners and residents, foregrounding participation, responsibility, and context rather than treating heritage as static "content" to be documented [3] [15]. We use interdisciplinary making to support its sustainability in the digital era: turning vulnerability into creative inquiry. From research to making, these processes lead to creative responses discussed in the following section.

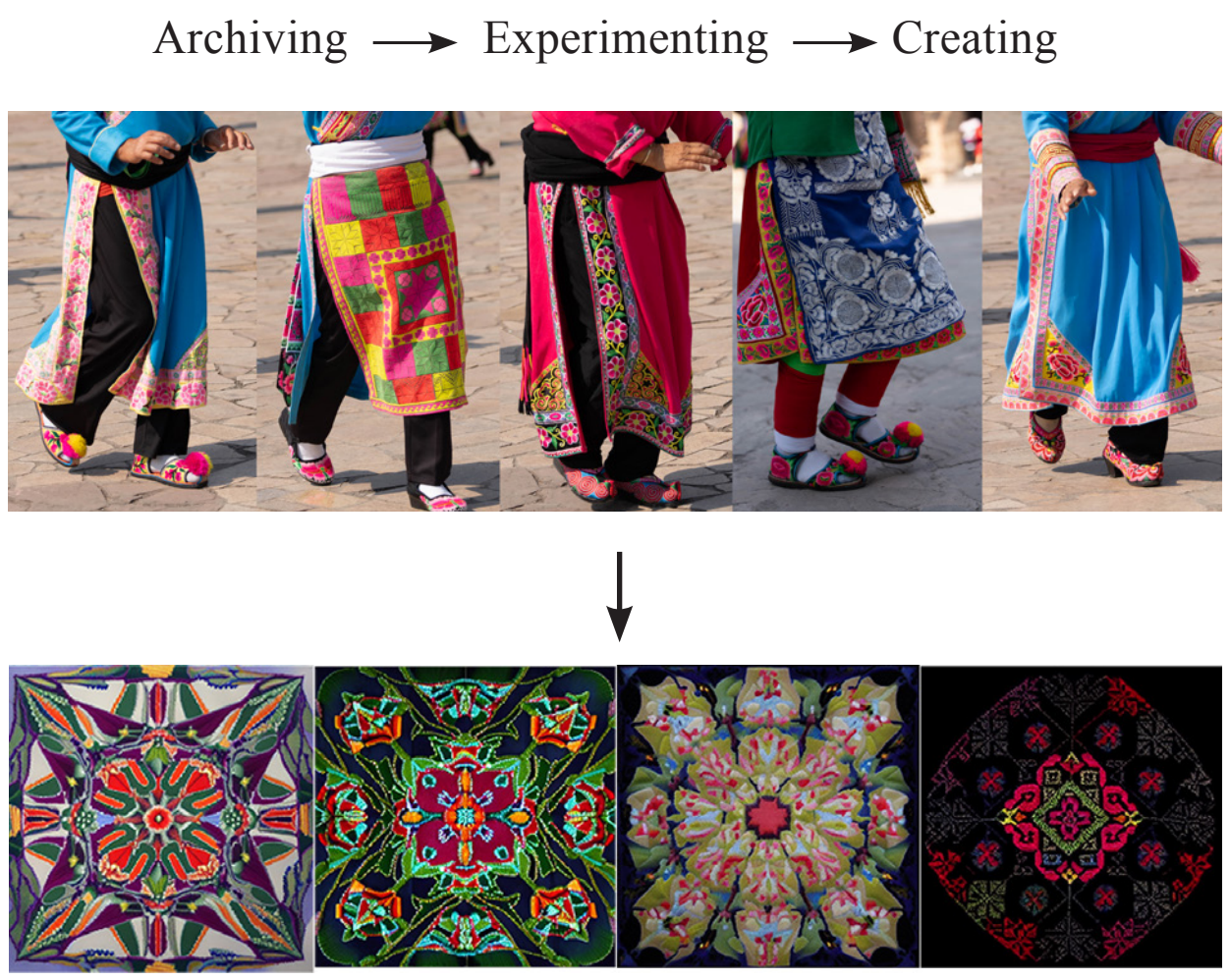

Fig 6: Cultural creativity implementation sample case

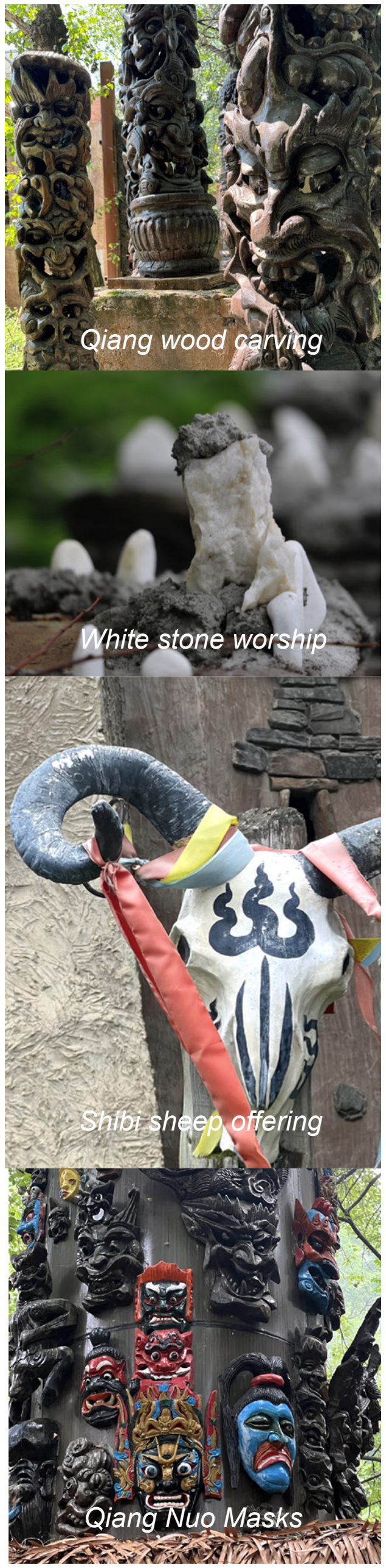

Fig 7: Worship system

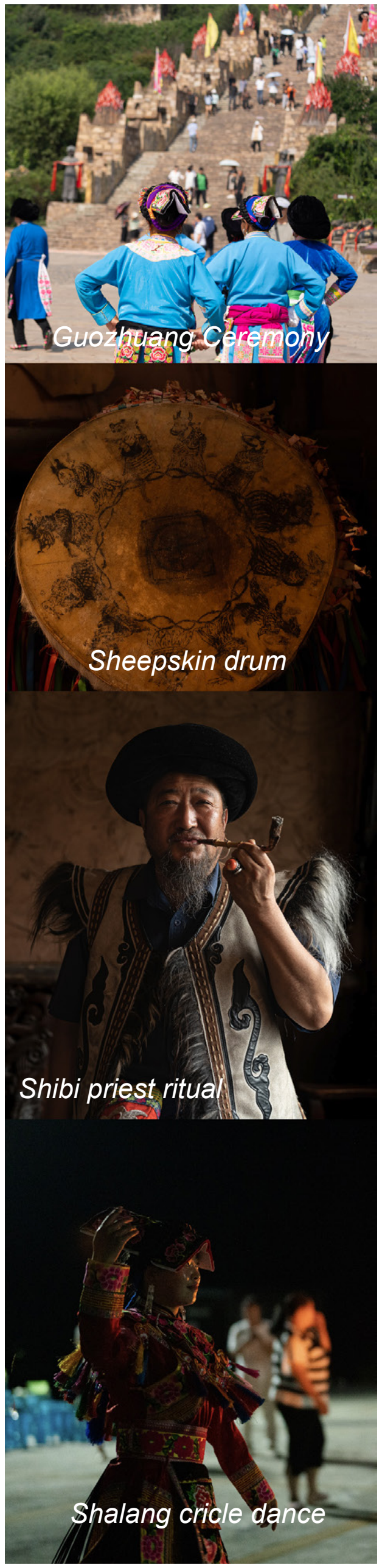

Fig 8: Dance and rituals

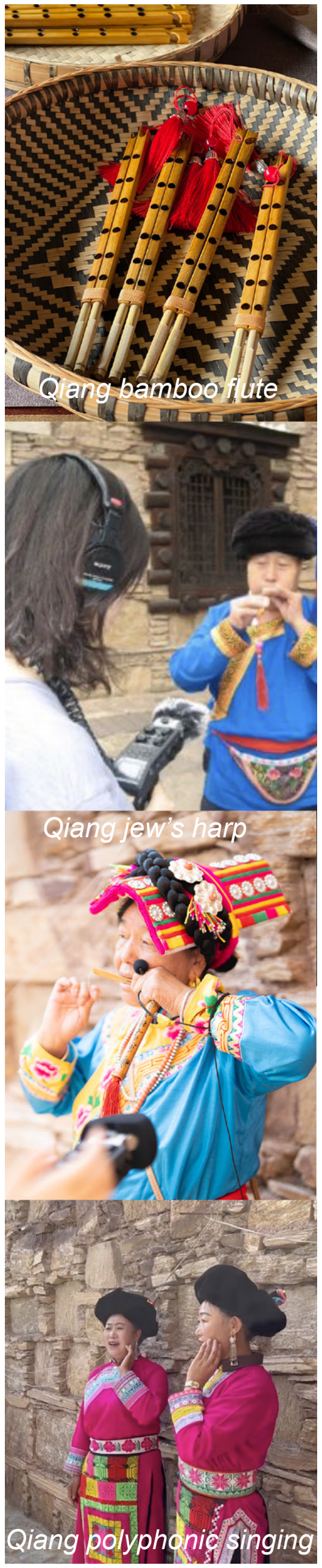

Fig 9: Qiang music

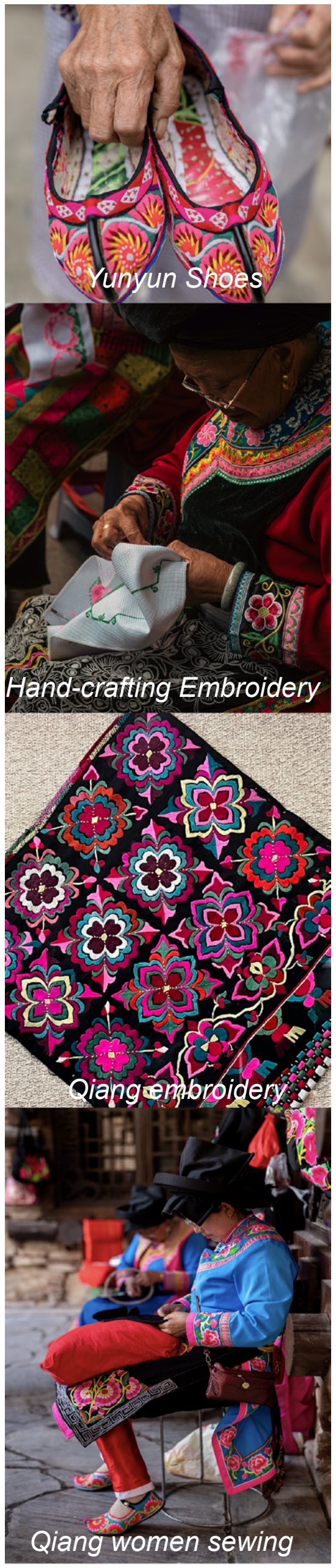

Fig 10: Qiang embroidery

## EMERGING CREATIVE MODES FROM RESIDENCY
### Case Study Series 1:
### Ambiguity for Being Part of Nature

In the context of the *Digital Qiang Garden* residency, creativity emerges from what Gaver et al. describe as ambiguity as a resource for design [16], where the lack of clear constraints compels artists to critique in sense-making and creativity-exploring through being with nature, which stems from the unpredictable agency of the terrain, the weather, and the more-than-human existence of the Qiang valley area as Figure 11.

*My Body Is Not My Shrine* (2023) is a case study for understanding creativity as an ecological attunement. Composed of lightweight and organic materials installed in the valley, its visibility and structure are entirely dependent on the fluctuating wind, fog, and humidity of the Aba Prefecture. From a phenomenological perspective, this work dissolves the boundary between the artist as an active subject and nature as a passive object. It's seeking to articulate a continuity between the body and the world, to embody what Merleau-Ponty terms *La chair du monde* (The Flesh of the World) [17], where the flesh is connective tissue that binds the seer and seen, while the body is not merely in the world but of the world. The artistic authorship is deliberately ceded to atmospheric essence, as a traditional, local, and, climatic event that grows from it. It aligns with Tim Ingold's anthropological assertion that making is a process of correspondence, a dance with the substance world rather than a dictation of form [18]. Ambiguity here is a method of inquiry to discover the boundary between art, culture, and environment.

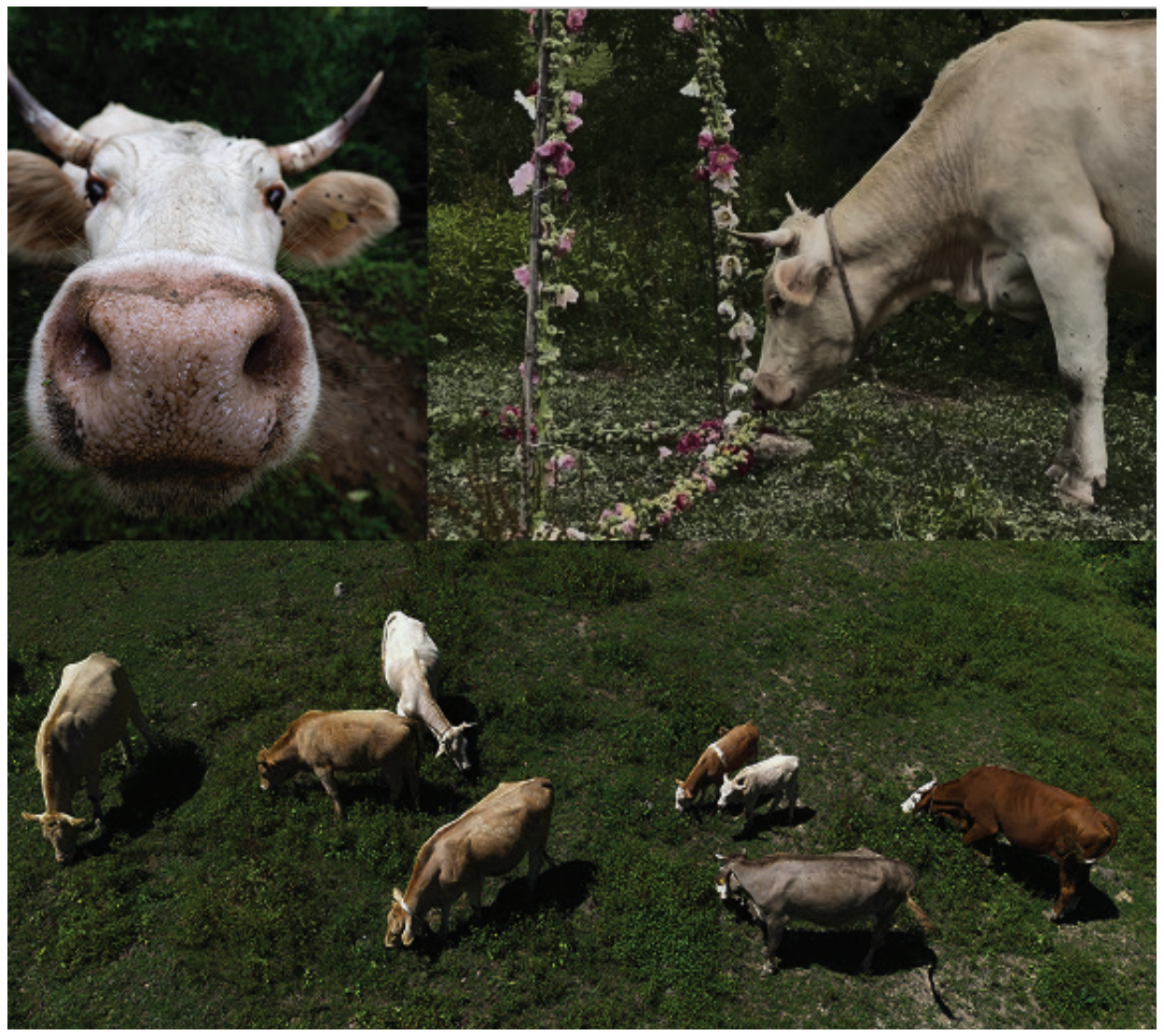

*Fig 11: Metabolic interaction process illustrates the artwork's fluid reintegration into the biological cycle*

Manifesting as a performance-installation hybird, the performance *My Body Is Not My Shrine* negotiates with space and atmosphere, mirroring how installation *The New Sun* (2023) engages with time and substance. Constructed from eight types of organic materials collected from local households and the surrounding ecosystem, it forms a ritualistic circle honoring the Qiang people's nature worship. Crucially, the work is designed to decay. When digital culture is driven by an obsession with cloud storage and the preservation of data, representing what Chun describes as the enduring ephemeral, a paradoxical desire to make degenerative digital signals permanent [19], the deliberate decomposition of *The New Sun* reflects with Kairos as the qualitative and opportune moment associated with natural cycles and seasons [20]. Inspiring from Parikka's geology of media [21], the work acknowledges that materials are active participants in a cycle of entropy. As documented in Figure 12, the boundary of the artwork remains porous, inviting non-human agents to intervene. Smoke signals visualize the invisible flow of wind, while the flower-framed structure acts as a porous membrane, situating the performing body within the fluctuating meteorology of the valley as Figure 14. It illustrates the transformation of the installation into what Gernot Böhme terms *Atmospheric Architecture* [22], and the structure here composed of delicate flower stems defines space through tuned atmospheres with openess.

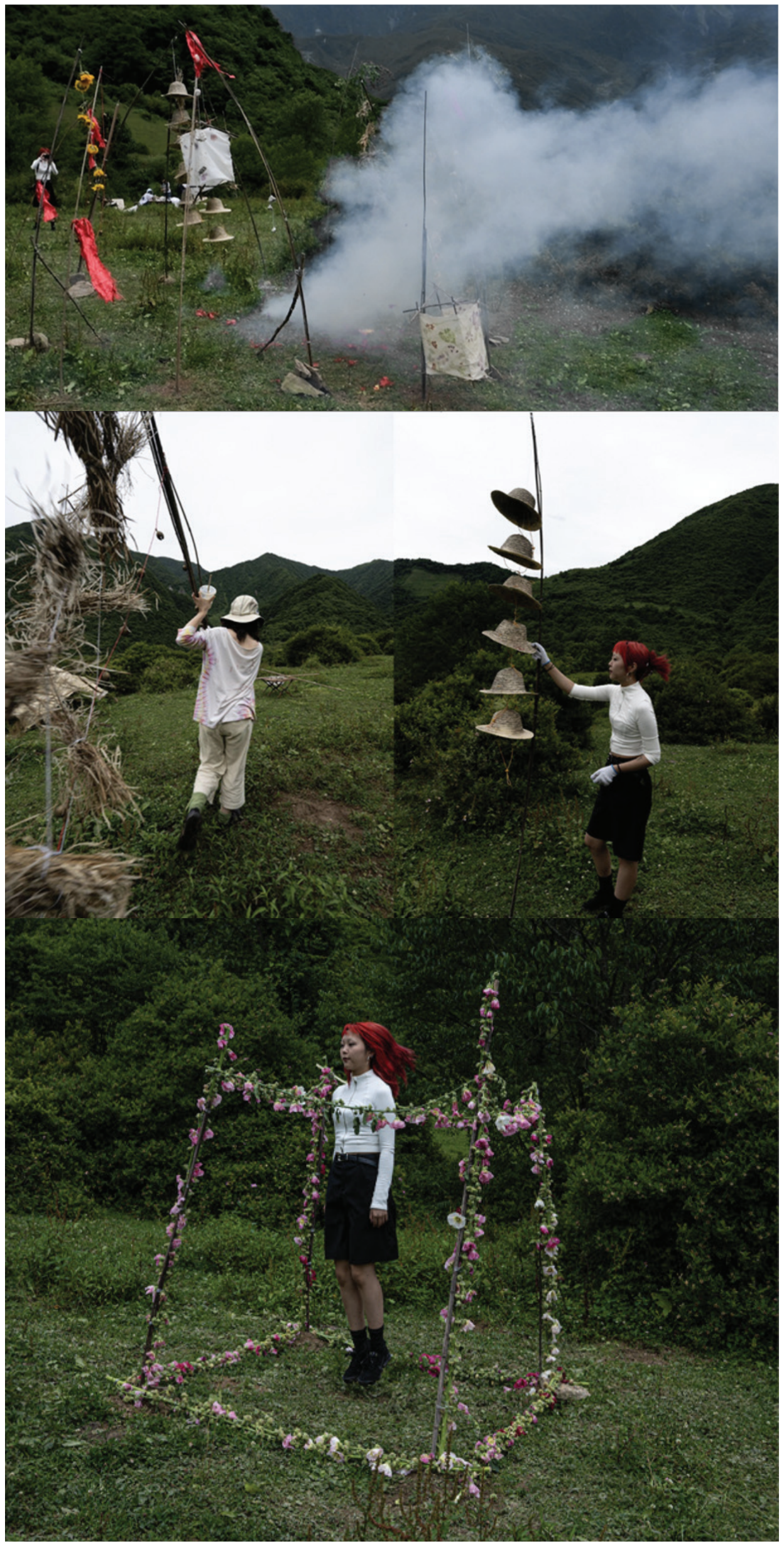

*Fig 12: Details of performance-installation hybrid flux*

## Case Study Series 2: Materiality as Mediating Immateriality

Moving beyond the hylomorphic tradition, where active human minds impose form upon passive matter, this section discusses the sensory anthropology and social semiotics perspective to analyze residency's outputs. We view the creative acts in the Qiang valley as a process of intra-action, a term coined by Karen Barad to describe the mutual constitution of subject and object [23]. Here, artists do not command the material; rather, intelligence emerges from the friction between the creator's intent and the material's affordances. From the biological decay of pigments to the sonic resonance of animal bone, the works analyzed below demonstrate that matter possesses its own agency, actively participating in the encoding and decoding of cultural memory.

Originating from a spontaneous interaction with the local community during initial oil pastel sketching sessions, The Roots (2024) evolved into a mural and installation. The final materiality was negotiated and co-decided with the residents, utilizing brushes fashioned from local bamboo poles and other raw materials from the environment. By replacing the neutral, white-cube gallery pedestal with the biomorphic matter motivated by the environment itself as Fig 13, *The Roots* reconstructed the frame that typically separates art from nature. It engages what Laura U. Marks defines as haptic visuality, a way of looking where the eyes function like organs of touch, moving over the surface rather than mastering it from a distance [24]. The medium here is the message: supporting structures are the living context of the image. It discovered that vision is not an abstract optical process but an embodied encounter, grounded in the specific geology and botany of the Qiang valley.

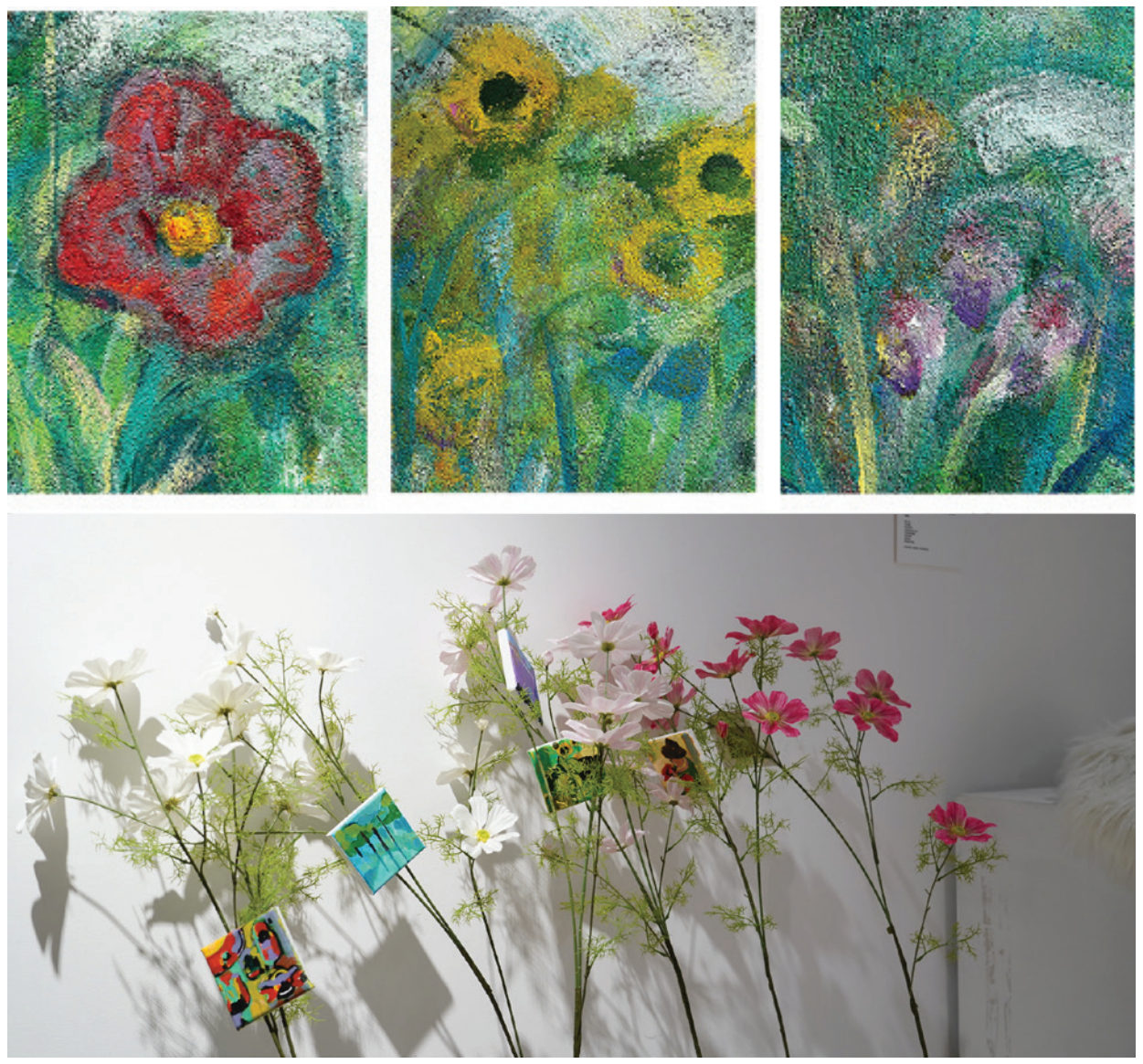

Fig 13: Art transformation from mural to installation

The Temp Tattoo (2024) project represents a possibility of shift in the semiotic circulation of heritage. By extracting traditional embroidery Qiang flower patterns, originally confined to textiles and garments, and re-purposing them as temporary tattoos as the Figure 14 shown, the project executes a process of embodied transcoding. In semiotic terms, it aims to alter the modal resources of the cultural sign from thread-on-fabric to ink-on-skin with multimodality [25], and it transforms the static heritage artifact into a skin interface as Gunther Kress mentioned [26]. Unlike exhibit that is observed, the tattoo is worn, performed, and eventually faded.

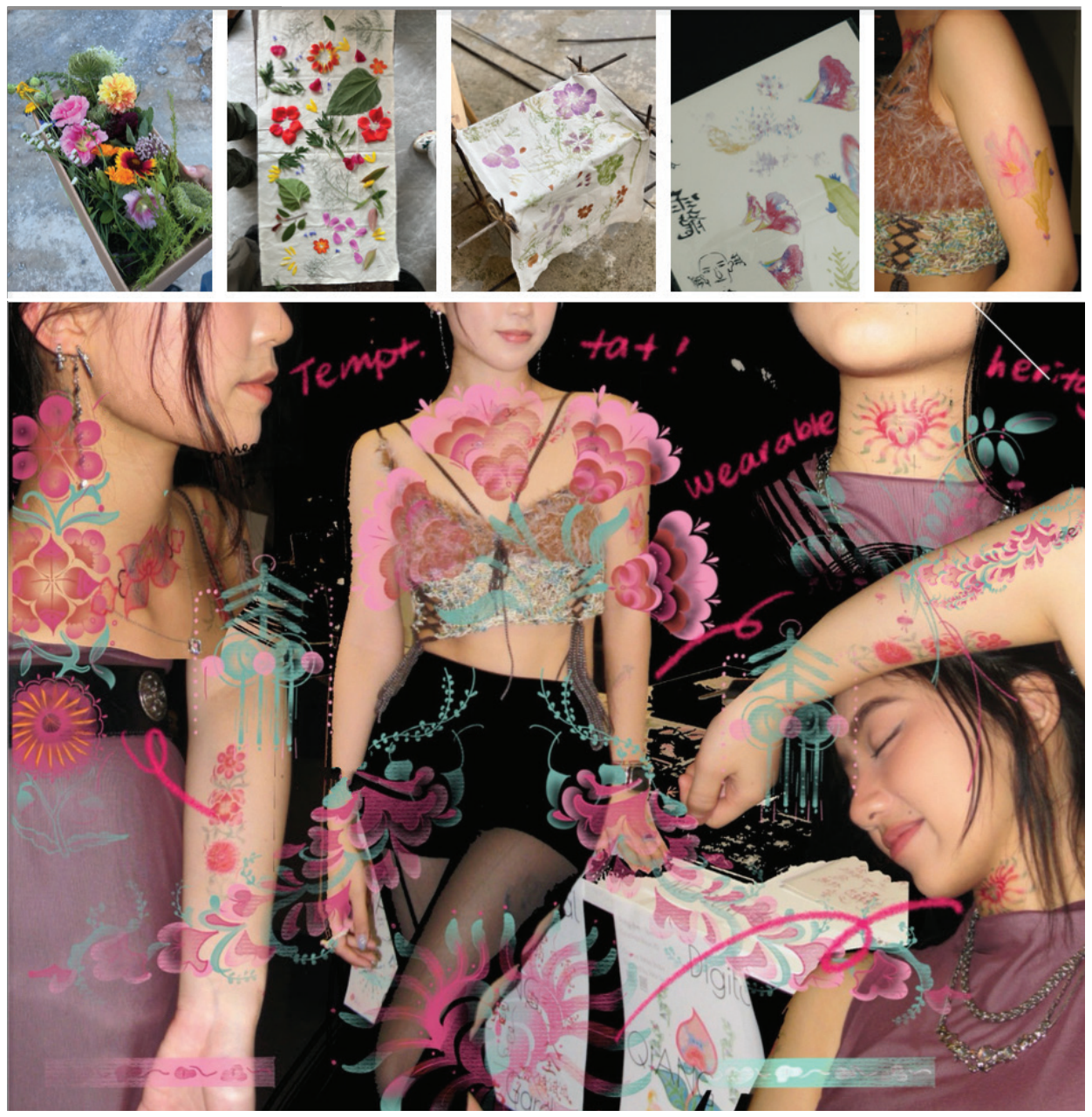

Fig 14: Iteration process from flower to tatoo design

The interactive installation The Sheep of Rrmea (2025) bridges the gap between ancient spiritual techniques and modern computational interfaces. By integrating sensors with a traditional sheepskin drum and animal bones, the work trans-mediates the Qiang people's most ancient communication device to communicate with deities, the shamanic drum, to a digital interface that triggers audiovisual feedback. Through the lens of media archaeology, specifically Siegfried Zielinski's concept of the *Deep Time of the Media* [27], we analyze the sheepskin drum as a sophisticated ritual medium designed for transmission across ontological realms. It connects ritual media (the bone, the skin, the beat) with computational media (the code, the sensor, the pixel), by performing a techno-spiritual translation as Figure 15.

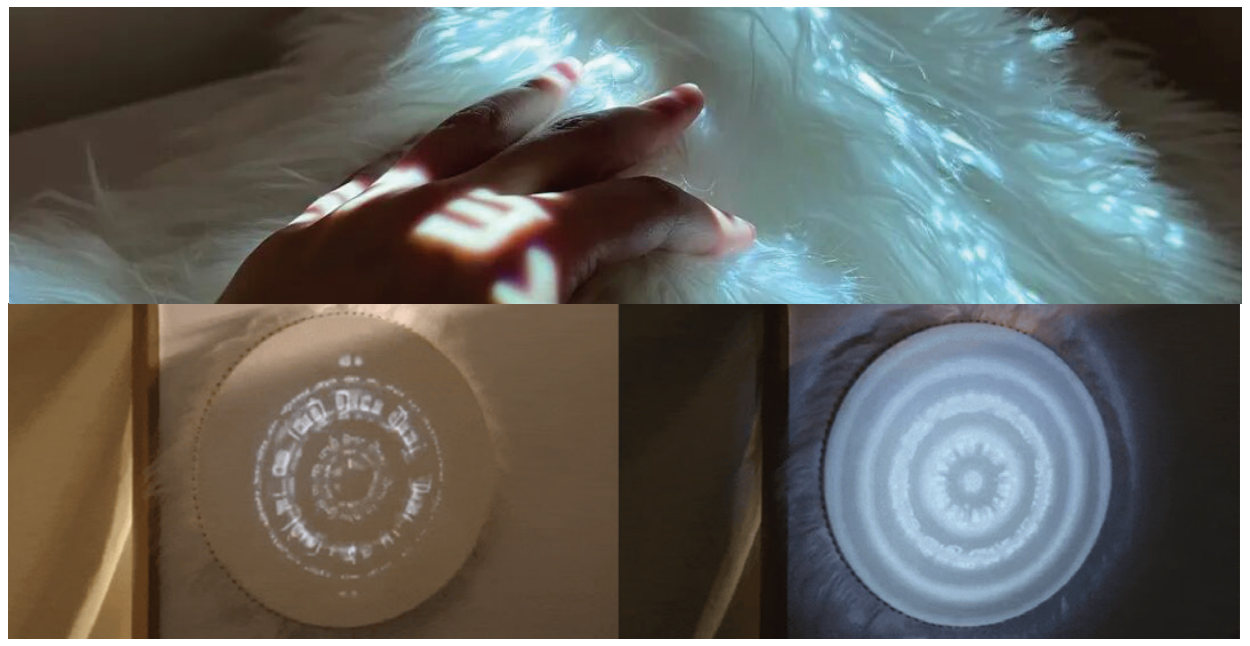

Fig 15: Multi-sensation of ritual media to digital media

**Case Study Series 3: Collective Authorship for Weaving Narratives**

*Digital Qiang Garden* residency operates on a model of distributed authorship, where creativity is emergent from a network of actors as human, algorithmic, and communal. Drawing on Edwin Hutchins' framework of distributed cognition [28], we posit that the cognitive system of the residency extends beyond the artists' bodies to include computational agents, local students, and village children. It further resonates with Hiroshi Ishii's Hiroshi Ishii's vision of *Tele Absence*, which extends the concept of telepresence to bridge the past and afterlife, thereby allowing the collective narrative to weave together voices across disparate times and spaces [29]. The following works illustrate how authorship is redistributed across the human-machine loop and the social contact zone with audience.

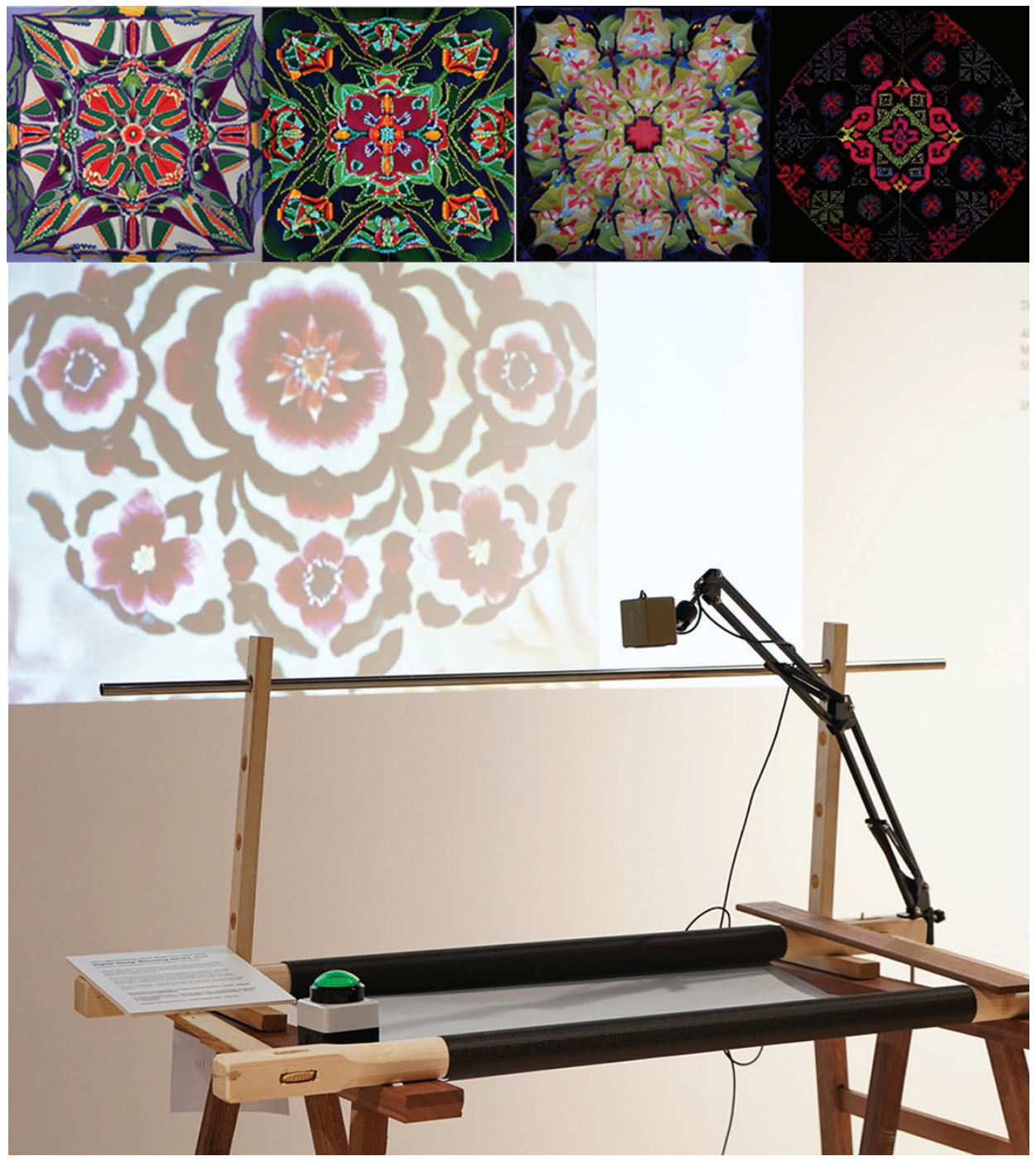

Fig 16: Generative patterns with real-time interaction

The AI interactive installation *Blooming Stroke* (2025) in the Figure 16 reconfigures the role of the designer. In this work, the visual output, intricate variations of Qiang floral patterns, is not pre-rendered by the artists but dynamically generated through a real-time feedback loop between the audience's input and a machine learning algorithm trained on a database of traditional embroidery. From the perspective of Second-order Cybernetics, the artist here moves from being an author to a designer of constraints. The AI system functions not merely as a tool, but as a quasi-autonomous agent with its own computational agency [30]. Following N. Katherine Hayles' analysis of the posthuman condition, subjectivity is no longer the exclusive province of the biological human but is constituted through the interface with intelligent machines [31]. In Blooming Stroke, the authorship becomes algorithmic and participatory. The AI acts as a co-creator, generating infinite variations of the Qiang flower that no single human designer could conceive or execute manually. This creates a human-machine loop where the cultural heritage database acts as the memory, the algorithm acts as the processor, and the audience acts as the trigger.

While *Blooming Stroke* distributes authorship to the machine, the collaborative fiber art project *Cloud Melody of Rrmea* (2024) and the Children's Workshops in Page 4 distribute it horizontally across the social spectrum. *Cloud Melody* involves co-creation between academy-trained artists and students, while the workshops invite village children to reinterpret their own visual culture. This process can be analyzed through Mikhail Bakhtin's literary theory of polyphony [32], that the residency functions as a dialogic space where no single voice holds authoritative dominance. Instead, meaning is generated through the friction and interaction between these voices. In Social Semiotics, this represents the socialization of signs [33]. The traditional Qiang symbols, originally rigid semantic units within the ritual or domestic sphere, are deconstructed by the artists and then reconstructed by the children. As observed in the workshops, children blend traditional motifs with personal imagination, creating a vernacular modernism. Furthermore, referencing Mary Louise Pratt's concept of the contact zone [34], the residency serves as a social space where cultures meet, clash, and grapple with each other. The resulting artworks as in Figure 17 are *Boundary Objects* [35], that facilitate communication between the tacit knowledge of the local community and the explicit knowledge of the visiting researchers.

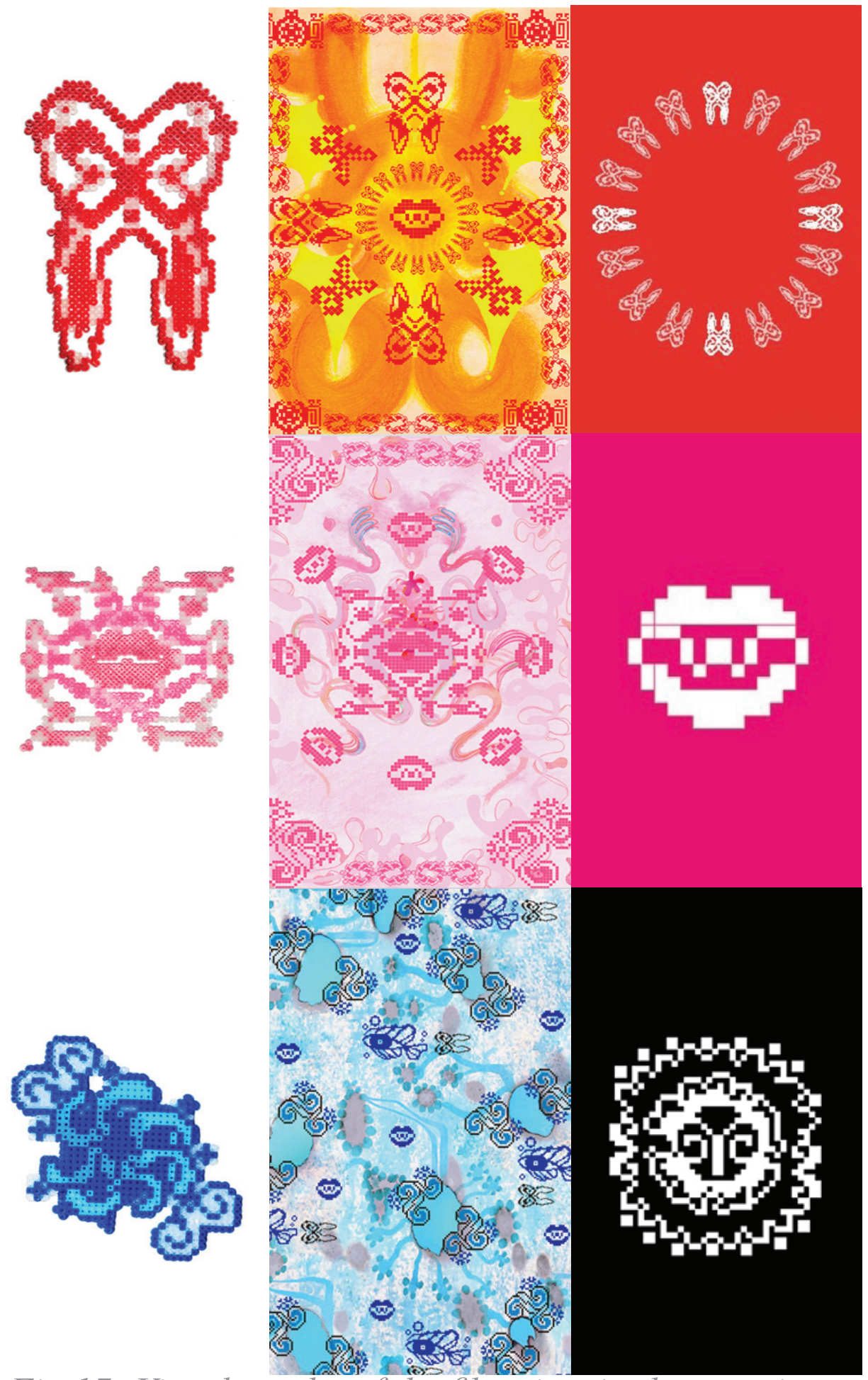

Fig 17: Visual results of the fiber-inspired art project as collective creation for community's narratives

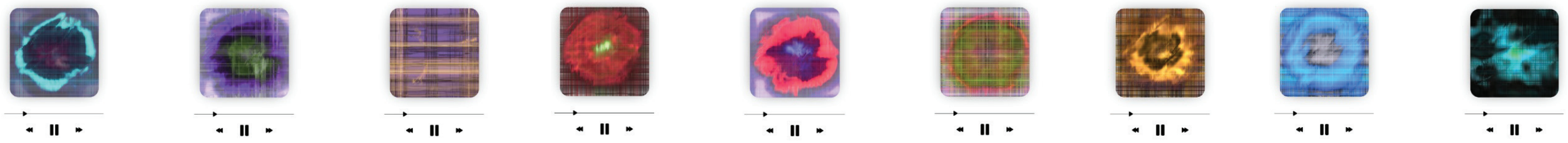

## Case Study Series 4: Continuity Beyond the Residency

As a transferable interface, the art residency acts as an emerging agent that negotiates ongoing encounters, from exhibitions to new sites, ensuring creative continuity through sustained reciprocity with local residents. This phase addresses the challenge of maintaining cultural vitality when physical co-presence ceases. By transforming heritage into hypermedia archives, the works demonstrate how residency-based creativity can deeply engage with living practices and local economies rather than observing them from a distance. Drawing on Hiroshi Ishii's vision of TeleAbsence and Walter Ong's theories of orality, these projects foster a trans-spatial presence that reactivates the village's aesthetic logic as a shared resource, ensuring that social impact and creative momentum persist long after the artists depart.

The *AR Digital Qiang Garden* (2024) application serves as a technological and cultural mediation that displaces the spatial experience of the Qiang valley into remote urban environments as New York, Boston, and Shanghai.

*Fig 19: As an autonomous interface, artwork Music ID Card feeds the archival body to perpetuate creativity*

By overlaying 3D reconstruction results of embroidery patterns onto city streets, the work creates a hybrid reality. The process can be analyzed through Hiroshi Ishii's framework of tangible bits, specifically his vision of seamlessly coupling the dual worlds of bits and atoms [36]. While Ishii's *Radical Atoms* vision focuses on dynamic physical materials, the core philosophy remains to liberate digital information from the confinement of screens and manifest it in the physical environment. The AR application here acts as a tele-presence interface, allowing the specific genius loci (spirit of place), a concept articulated by Christian Norberg-Schulz [37]. This constitutes a form of spatial displacement. The AR garden implants the village's aesthetic essence onto the urban fabric and enables a *Peripheral Awareness* as Hiroshi Ishii describes [38], turning the city itself into a canvas for indigenous narrative, thereby dissolving the binary between the remote rural and the immediate urban as Figure 18 shown.

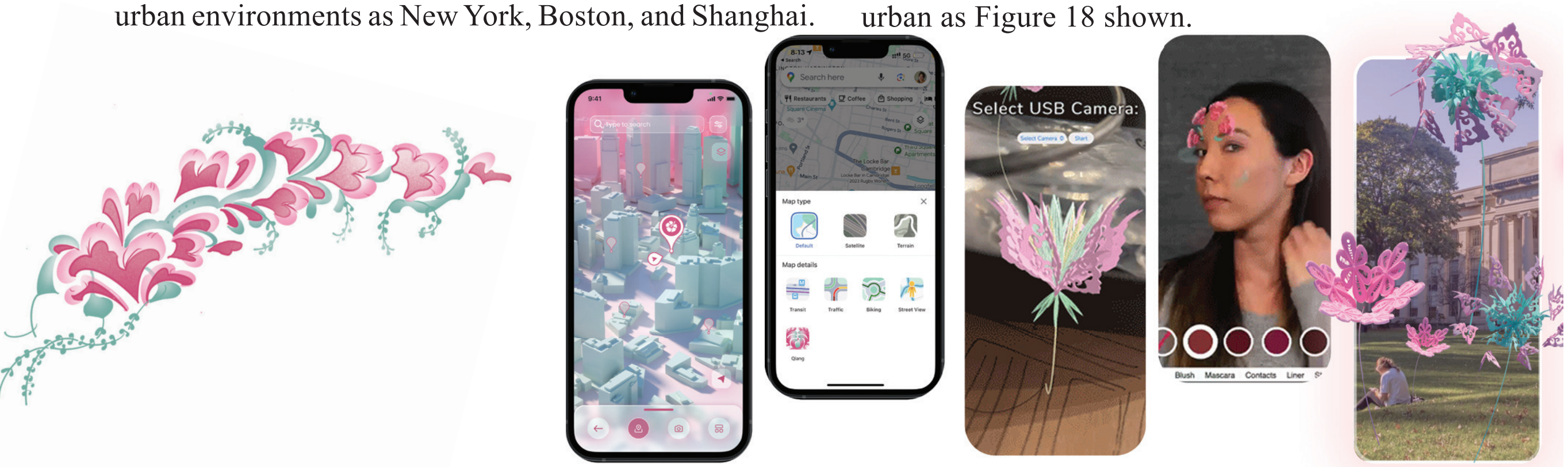

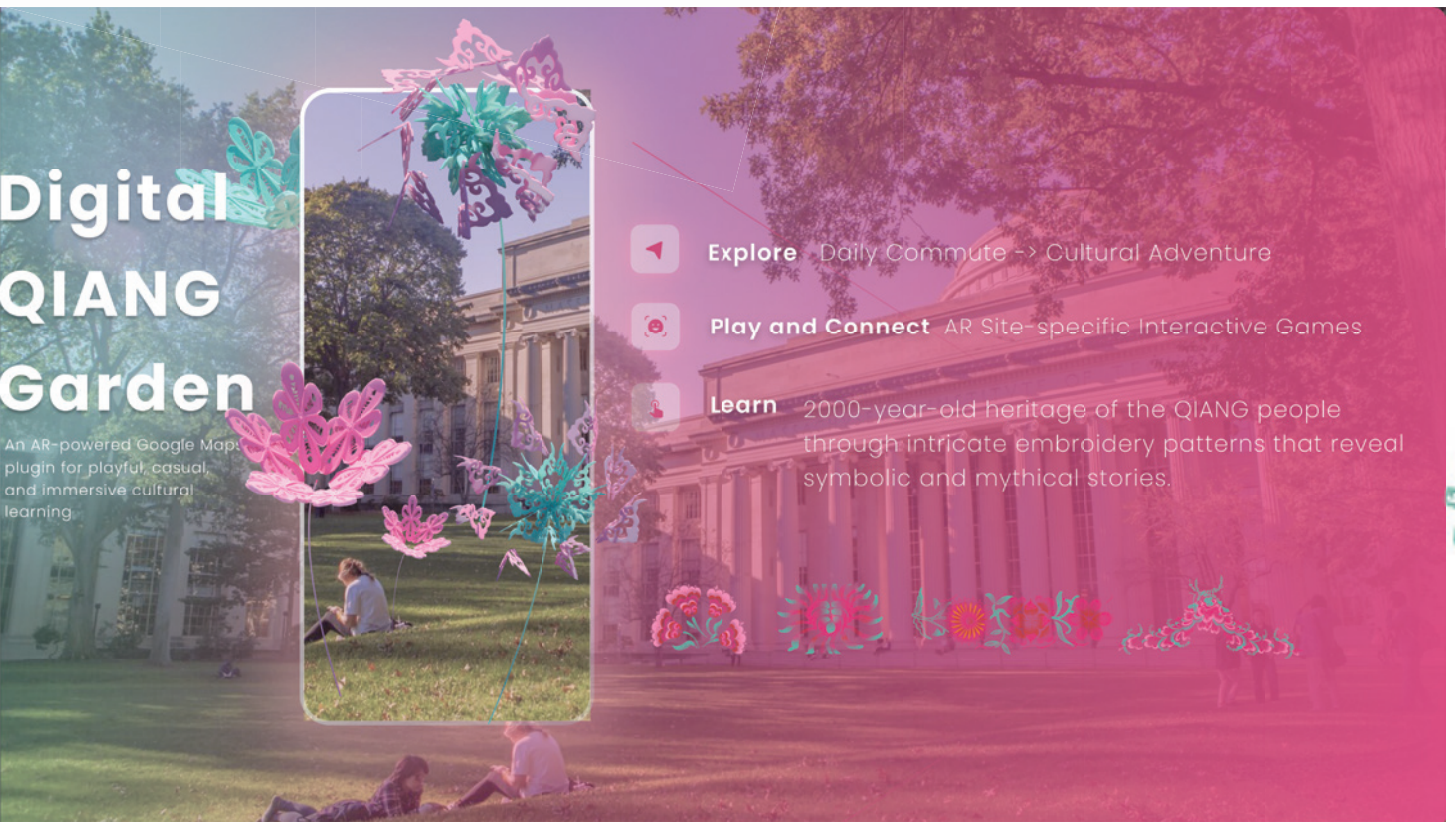

*Fig 18: The AR interface functions as a technological mediation that transforms the city into a canvas for Qiang aesthetics by implanting indigenous narratives*

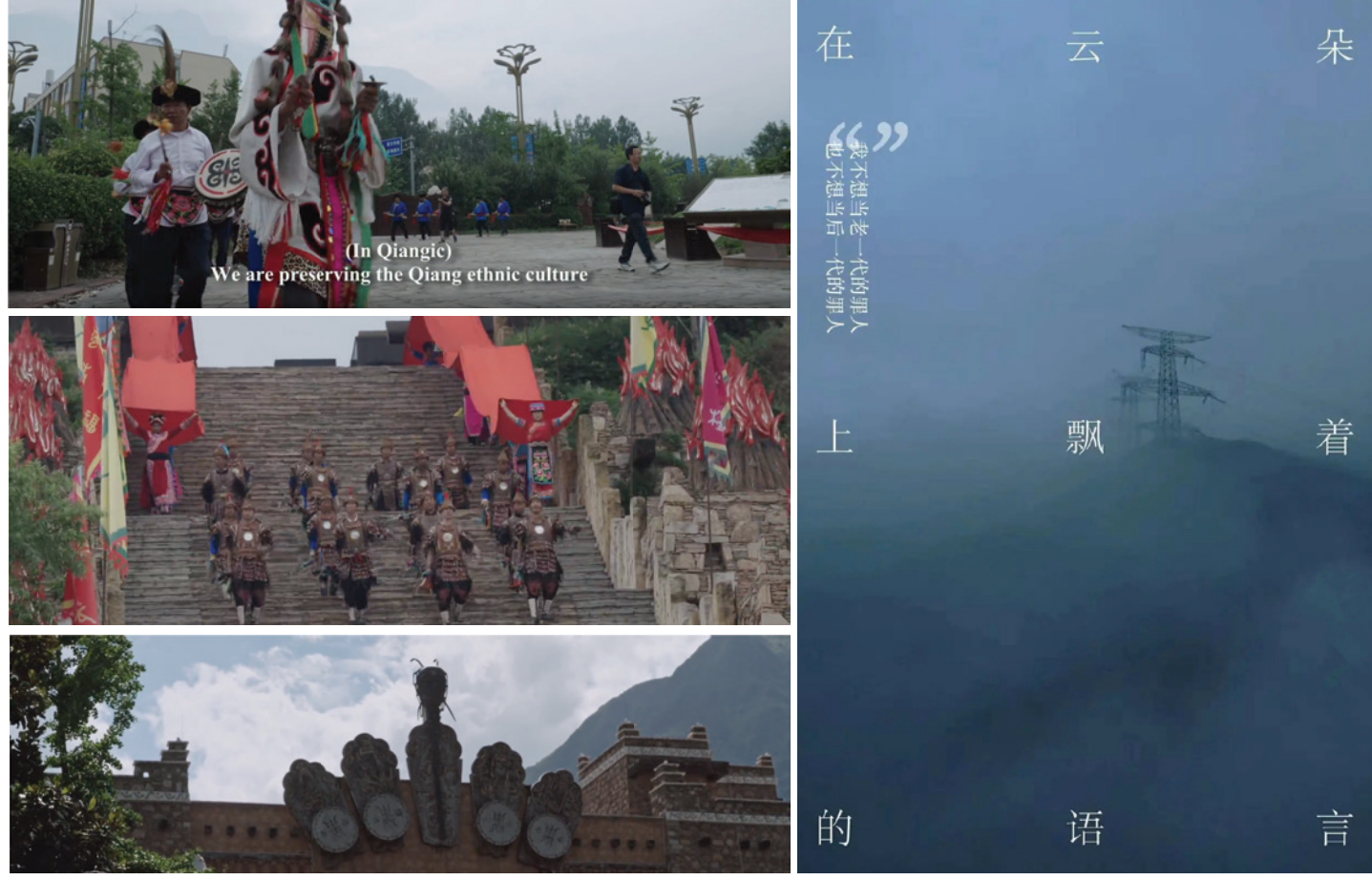

*Fig 20: Poster and key frames of the documentary*

The *Digital Qiang Garden Music ID Card* (2024) as Figure 19 and the documentary *The Language Floating* (2024) as Figure 20 address the ontological crisis of Qiang culture: the lack of a written script spreading. The Qiang language and music exist purely in the realm of secondary orality, vulnerable to erasure in a text-dominant modernization process. *The Music ID Card*, which visualizes sound waves of traditional instruments by the Qiang flute and sheepskin drum into graphic identity cards, performs an intersemiotic translation [39]. Drawing on Roman Jakobson's semiotic theory, the work transmutes signs from the auditory system into the visual system. Furthermore, the documentary The Language Floating facilitates a shift from orality to hypermedia. Referencing Walter Ong's distinction between the psychodynamics of oral and literate cultures [40], the project captures the ephemeral, event-based nature of oral storytelling and preserves it within a non-linear, accessible digital structure. By uploading these vanishing languages to the cloud, the project ensures their persistence as digital signals. However, unlike static museum archives, these hypermedia artifacts function according to the logic of heritage futures [41]: they are not preserved for the sake of the past but reactivated as living resources for future creation. The digital archive becomes a generative database from which new melodies and narratives can be retrieved and remixed, ensuring the culture survives through evolution rather than mere stasis.

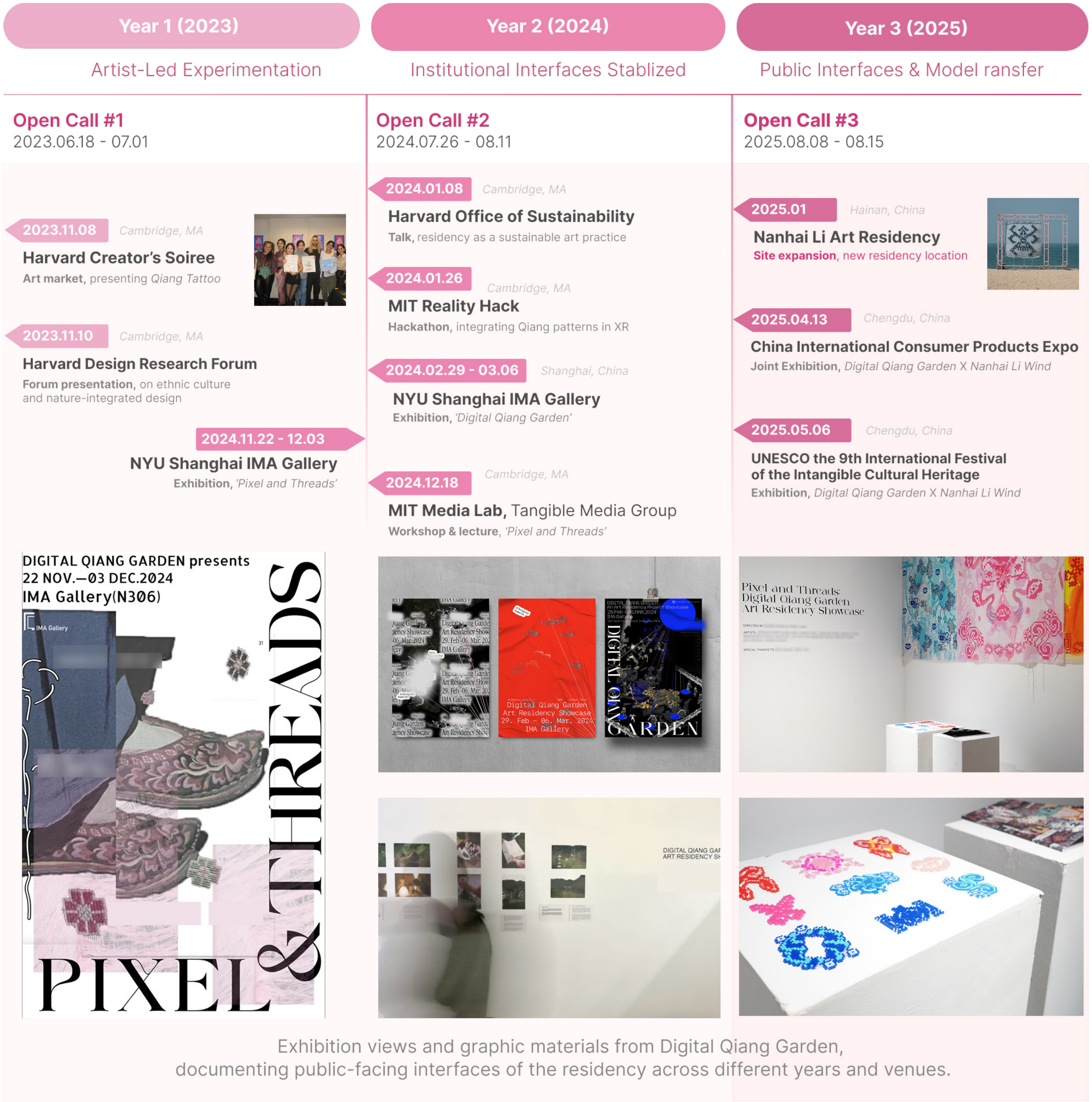

*Fig 21: Overview of the residency's impact through evolution*

### IMPACT, REACH, AND EVOLUTION

As shown in Figure 21, the residency functions as a transferable structure that evolved through repeated iterations, forming a rhythmic pattern of engagement. This recurring framework allows artistic practices and long-term relationships with permanent residents to accumulate, allowing social impact to remain rooted in local perspectives rather than extractive exhibition outcomes.

Across three years, this translation took multiple forms: two gallery exhibitions in Shanghai attracted over 2,000 on-site visitors, while two national-level exhibitions/expos reached an estimated 5,000 daily visitors, extending the residency's questions into broader public discourse. At the institutional level, five lectures, workshops, and exhibitions were hosted across departments at universities including Harvard and MIT, engaging students and researchers through shared pedagogical and archival inquiry. Online dissemination and exhibition documentation further expanded reach through social media circulation, allowing the work to travel without displacing its local grounding.

Accumulated impact from public exhibitions, institutional collaborations, and educational programs has generated trust, visibility, and relational capital for the residency, enabling it to function as a scalable interface. Through this momentum, the residency has been translated across diverse contexts, from heritage villages to classrooms and exhibition spaces, demonstrating its adaptability to different ethnic and economic environments. These translations show that residency-based creativity can engage deeply with living cultural practices, local economies, and everyday life rather than observing them from a distance, ensuring that culture is sustained through active, horizontal co-creation and mutual growth.

**Creativity for Change as a Layered Process**

*Qigital Qiang Garden* began as an impulse to act has evolved alongside broader changes in infrastructure, policy, and participants' lives. The residency as interface allows multiple forces of change to pass through, interact, and be reconfigured, approaching creativity for change as a layered and relational process.

Fgure 22 maps the layered changes that emerged through *Digital Qiang Garden* as an artist-led residency interface. At the organizational level, the residency reconfigured how artistic opportunities are produced by shifting away from institution-centered models toward collective input, shared resources, and open participation, enabling emerging and self-taught artists to work outside competitive hierarchies.

At the level of artistic practice, creation was shaped by prolonged immersion in place, where materials, narratives, and rhythms were encountered through daily living, fieldwork, and uncertainty. Cultural heritage engagement operated as a co-creative and investigative process, addressing the structural conditions affecting intangible heritage such as natural disasters, education systems, and development policies, and transforming vulnerability into contemporary artistic expression and living archives.

These conditions also influenced artists' life paths, as residency projects extended into thesis works, research directions, and long-term career shifts in how participants understood the social role of art.

Locally, the residency fostered reciprocal relationships with children, elders, heritage bearers, and small businesses through co-production and fair collaboration, while ecological and environmental concerns surfaced through situated attention to landscape, materials, and policy contexts.

As these layers accumulated, the residency generated practices capable of traveling outward through exhibitions, pedagogy, and cross-regional adaptations, forming the basis for the broader impacts and future transfers discussed in the following sections.

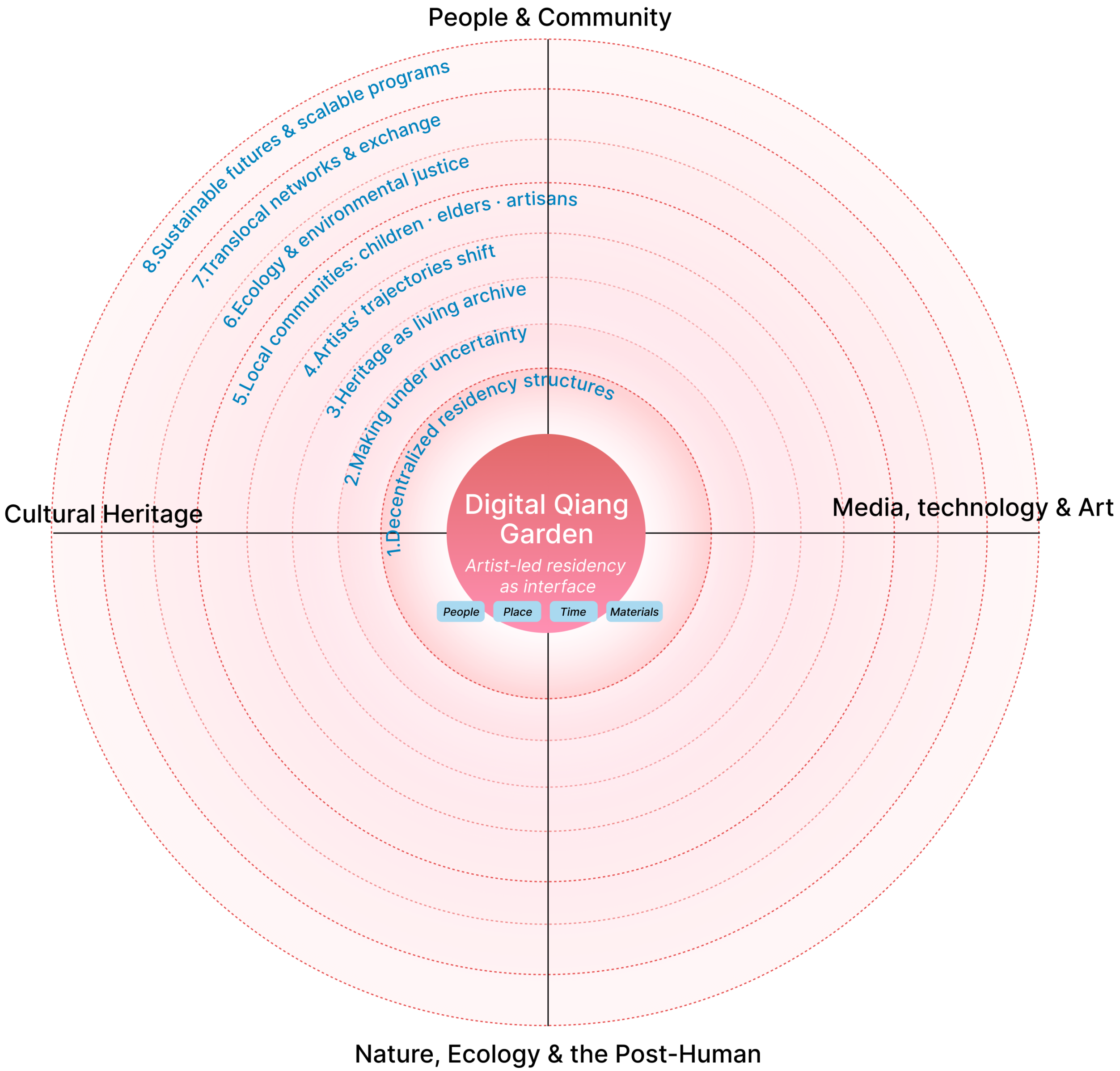

Fig 22: Diagram of eight layers of changes under foursocial and technological axis

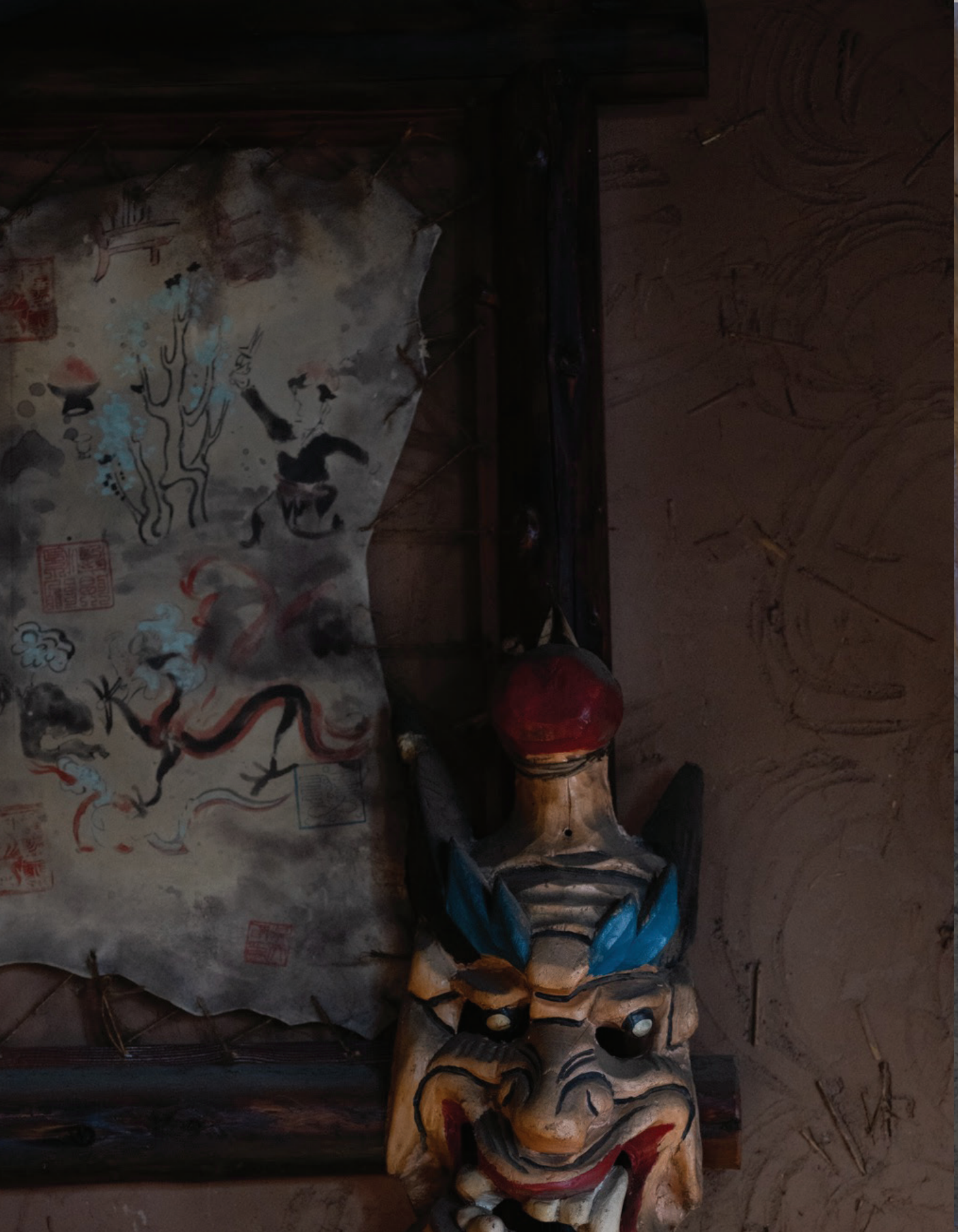

## REFLECTION

Reflecting practices, *Digital Qiang Garden* demonstrates how an artist-led residency functions as an evolving infrastructure that weaves through macro-level socio-economic shifts. By situating interdisciplinary artists within the village, the residency acts as a qualitative catalyst, transforming abstract development policies into tangible cultural capital and reshaping heritage preservation from static archiving to active re-creation. This initiative not only offers a transferable residency as methodology for rural revitalization but also fundamentally redefines artist mobility from individual competition to collective, distributed agency. Ultimately, this work contributes to HCI by presenting a scalable model of creativity under and for change, shifting the design focus from discrete tools to the social and material conditions that foster resilient, situated communities.

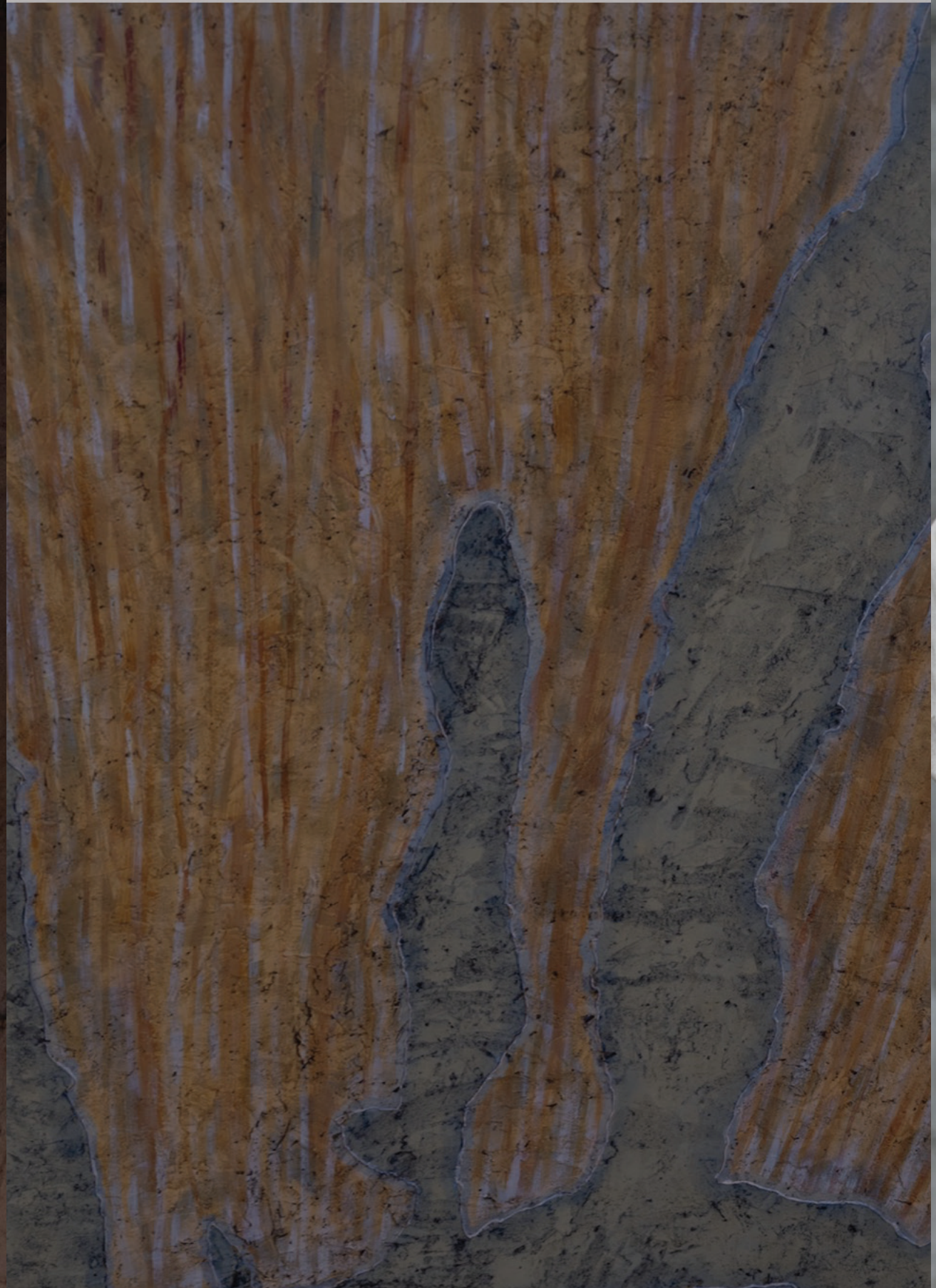

## FUTURE WORK

Moving forward, our vision transcends the single residency, aiming to establish a self-sustainable and scalable ecosystem for situated creativity. We are actively expanding this model to new ethnic contexts, with the Nanhai Li Art Residency launching in 2025 and further sites planned for 2026. The next phase prioritizes a deeper integration with local industries, specifically bridging artistic production with agriculture and cultural tourism, to generate tangible economic revenue for residents, thereby providing a solid economic foundation for the transmission of intangible cultural heritage. Through sustained international media exposure and deepening academic partnerships with higher education institutions, we aim to not only amplify community voices but also influence policy-making, demonstrating how grassroots artistic intervention can serve as a durable engine for rural revitalization.

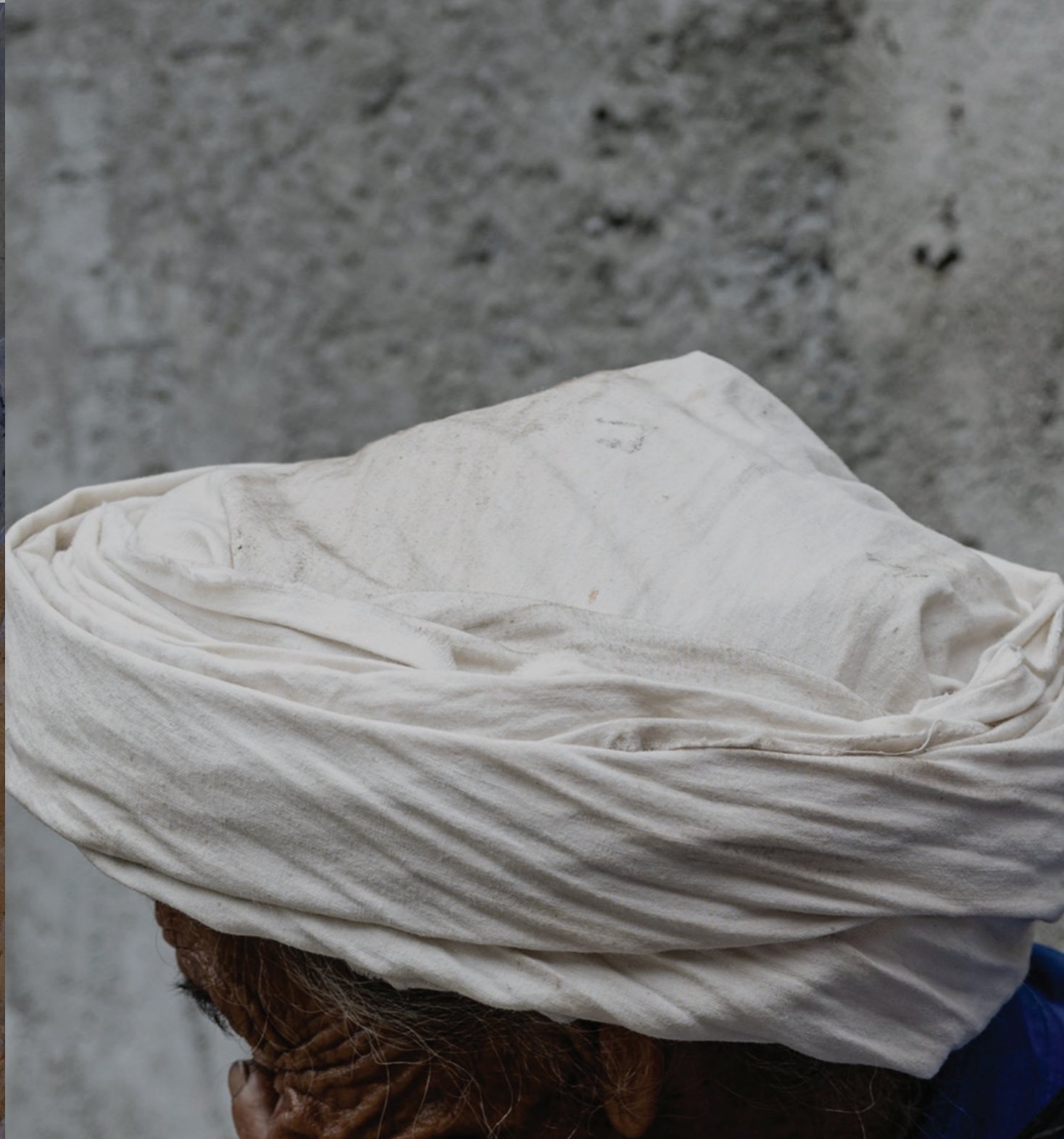

## CONCLUSION

This pictorial has presented *Digital Qiang Garden* as a practice-based inquiry into how an artist-led residency can function as a generative interface for creativity under change. By situating interdisciplinary practice within the specific ecological and cultural conditions of Qiang villages, we demonstrated that creativity emerges not through control, but through attunement to ambiguity, material engagement, and collective authorship. The residency model outlined here moves beyond the production of discrete artifacts to the design of relational infrastructures, conditions that allow tacit knowledge, heritage practices, and new technologies to intersect and evolve. Through our analysis of the residency's modes of praxis and its cascading impacts, we argue that artist-led initiatives offer a vital, transferable framework for navigating the complexities of cultural preservation and rural revitalization. Ultimately, *Digital Qiang Garden* contributes to HCI and design research by illustrating how situated, embodied, and collaborative making can weave new forms of resilience, ensuring that creativity remains a living, adaptive force in a changing world.